\documentclass[aps,twocolumn,prb,reprint,amsmath,amssymb,showpacs,superscriptaddress	,longbibliography]{revtex4-2}

\usepackage{graphicx}
\usepackage{dcolumn}% Align table columns on decimal point
\usepackage{bm}% bold math
\usepackage{amsmath}
\usepackage{color,xcolor}
\usepackage{mathrsfs}
\usepackage{float}
\usepackage{txfonts}
\usepackage{wasysym}
\usepackage{braket}

\newcommand{\JS}[1]{\textcolor{black}{#1}}

\makeatletter

\newcommand{\Rmnum}[1]{\expandafter\@slowromancap\romannumeral #1@}
\makeatother
\begin{document}
\title{Binding zero modes with fluxons in Josephson junctions of  time-reversal invariant topological superconductors} 

\author{Gabriel F. Rodr\'{\i}guez Ruiz}
\affiliation{Escuela de Ciencia y Tecnolog\'{\i}a  and ICIFI, Universidad Nacional de San Mart\'{\i}n-UNSAM, Av 25 de Mayo y Francia, 1650 Buenos Aires, Argentina}
\author{Adrian Reich}
\affiliation{Institute for Theory of Condensed Matter, Karlsruhe Institute of Technology (KIT), 76131 Karlsruhe, Germany }
\author{Alexander Shnirman}
\affiliation{Institute for Theory of Condensed Matter, Karlsruhe Institute of Technology (KIT), 76131 Karlsruhe, Germany }
\affiliation{Institute for Quantum Materials and Technologies, Karlsruhe Institute of Technology (KIT), 76131 Karlsruhe, Germany }
\author{J{\"o}rg Schmalian}
\affiliation{Institute for Theory of Condensed Matter, Karlsruhe Institute of Technology (KIT), 76131 Karlsruhe, Germany }
\affiliation{Institute for Quantum Materials and Technologies, Karlsruhe Institute of Technology (KIT), 76131 Karlsruhe, Germany }
\author{Liliana Arrachea}
\affiliation{Escuela de Ciencia y Tecnolog\'{\i}a  and ICIFI, Universidad Nacional de San Mart\'{\i}n-UNSAM, Av 25 de Mayo y Francia, 1650 Buenos Aires, Argentina}
\affiliation{Centro At\'omico Bariloche and Instituto de Nanociencia y Nanotecnolog\'{\i}a CONICET-CNEA (8400), San Carlos de Bariloche, Argentina}

\begin{abstract}
We study the joint dynamics of the phase bias $\phi$ and the propagating Majorana fermions of the edge modes in  Josephson junctions containing 2D time-reversal invariant topological superconductors (TRITOPS). We consider TRITOPS-TRITOPS junctions, as well as junctions between topological and non-topological superconductors (TRITOPS-S). 
Both types of junctions are described by effective Dirac Hamiltonians with a $\phi$-dependent mass. 
We analyze the effect of the  phase fluctuations in the junction, as well as 
solitonic solutions of $\phi$ generated by fluxons trapped in the junction. We show that these solitons generate a spatial-dependent mass with a sign change akin to the Jackiw-Rebbi model. This enables the formation of zero-energy fermionic states localized at the fluxon. For the TRITOPS-TRITOPS junction 
these consist of a Kramers pair of Majorana modes, while for the TRITOPS-S one, there is a single Majorana fermion. The localized  bound states hybridize in soliton-antisoliton configurations. Depending on the occupation state, these modes generate an effective attraction or repulsion in the dynamics of the soliton-antisoliton collision. 
\end{abstract}

\date{\today}
\maketitle

\section{Introduction} 
The search for  Majorana zero modes is one of the most active avenues of research in condensed matter physics. This is motivated by 
the relevance of topological properties of quantum matter \cite{bernevig2013topological, wen2017colloquium} and because of 
the potential use of these quasiparticles as building blocks of
fault-tolerant quantum computation \cite{kitaev2003fault,nayak2008non}. 
One-dimensional (1D) topological superconductors host these quasiparticles localized at the ends. 
Several platforms  to realize this phase have been proposed, including quantum wires \cite{wires1,wires2,mourik2012signatures,rokhinson2012fractional,das2012zero,albrecht2016exponential,deng2012majorana}, magnetic impurities \cite{yazdani,wiesendanger,ruby}, non-topological superconductors in proximity to topological insulators \cite{fu2009josephson} and  
 planar Josephson junctions controlled by the phase bias \cite{Planarwire,PplanarX,Planarfrac,haim2019benefits,hart2017controlled,ren2019topological,fornieri2019evidence}. In two-dimensional (2D) systems
 the topological superconductors host chiral Majorana modes propagating along the edges. Promising signatures of these modes have been observed in iron-based compounds \cite{zhang2018observation,wang2020evidence}
and van der Waals heterostructures \cite{kezilebieke2020topological}. These and other results have been  reviewed in Refs.~\onlinecite{alicea2012new,sato2017topological,aguado2017majorana,flensberg2021engineered}.

A different class of topological superconductors exist that preserve time-reversal symmetry. These belong to the
 DIII-class defined in Ref.~\onlinecite{ryu2010topological} and are often referred to as TRITOPS (time-reversal invariant topological superconductors) \cite{haim2019time}. 
 There are many proposals to realize this topological phase in 1D and 2D systems
 \cite{kwon,dumitrescu2013topological,haim2014time,tanaka-tritops,qi2009time,fu2010odd,scheurer2015topological,wong2012majorana,zhang2013time,keselman2013inducing,haim2016interaction,ebisu2016theory,reeg2017diii,santos2010superconductivity,klinovaja2014kramers,mellars2016signatures,parhizgar2017highly,casas2019proximity,iron}. The edge states of these topological superconductors appear in Kramers pairs and give rise to interesting physical properties like  fractional spin densities and unusual features in the Josephson effect 
\cite{keselman2013inducing,chung2013time,nakosai2013majorana,schrade2015proximity,li2016detection,camjayi2017fractional,schrade2018parity,
aligia2018entangled,gong2016influence,mashkoori2019impact,lauke2018friedel,arrachea2019catalog,
haim2019spontaneous,knapp2020fragility,francica2020topological,volpez2020time,chinellato2024topological,PhysRevB.109.075416}.

\begin{figure}[t]
\centering
\includegraphics[width=\linewidth, trim={0cm 6.5cm 0cm 2cm},clip]{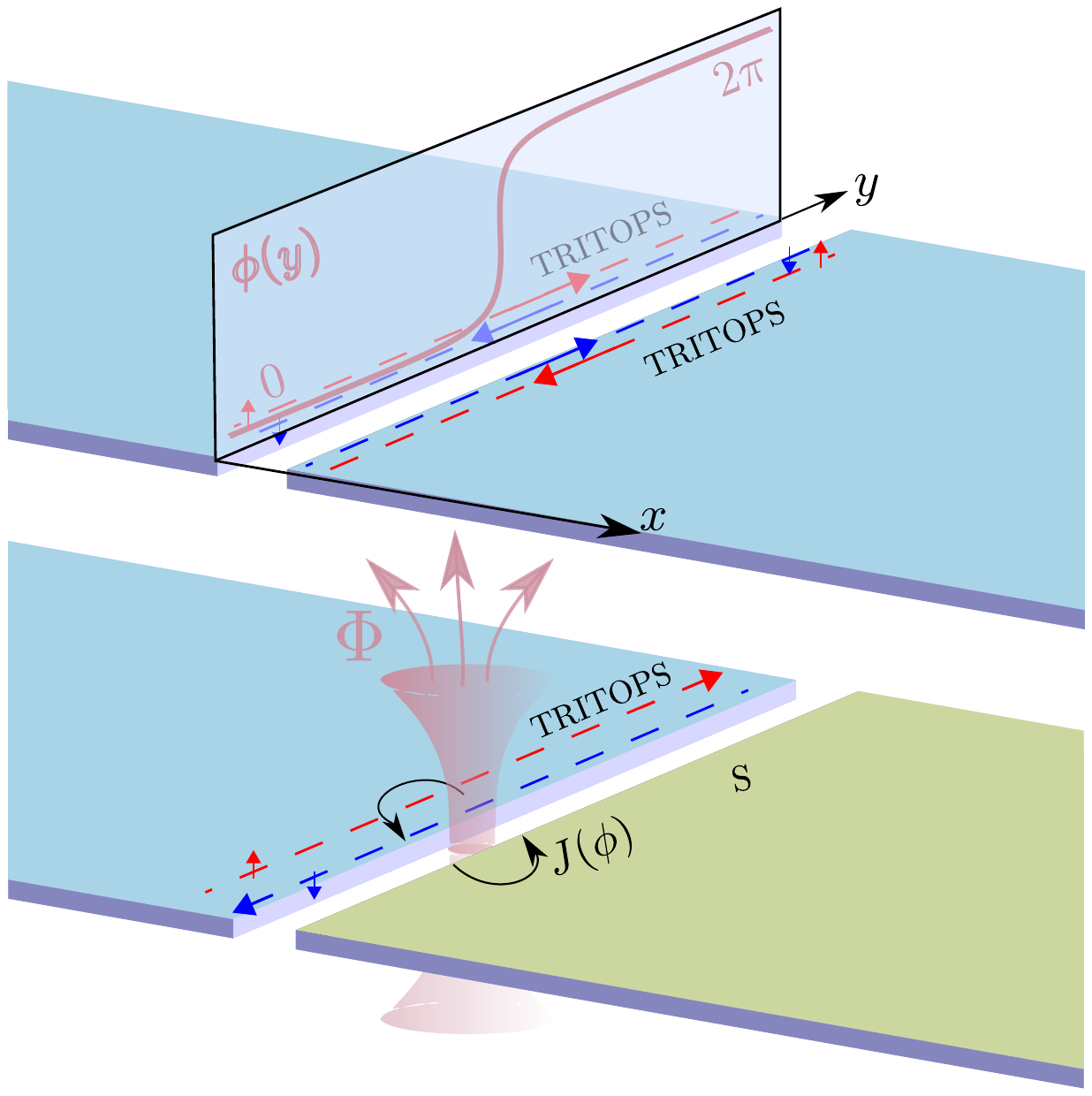}
\caption{Sketch of a TRITOPS-TRITOPS junction (top) with a soliton profile for the phase $\phi=2 \pi/\Phi_0$ and a TRITOPS-S junction (bottom) with a fluxon with flux $\Phi$. }
\label{fig:sketch}
\end{figure}

Recently, Josephson junctions with two-dimensional TRITOPS were studied in Ref.~\onlinecite{us}. The edge states of 2D-TRITOPS are 
 Kramers pairs of counter-propagating helical Majorana modes. These modes can be effectively described by a Dirac Hamiltonian of neutral fermions, i.e. a Majorana Hamiltonian, where the hybridization at the Josephson junction defines, \JS{depending on the nature of the junction, up to two masses  $m_1(\phi)$ and  $m_2(\phi)$. These mass terms depend} on the phase bias  $\phi$. Interestingly, when the junction is formed between a  TRITOPS and a time-reversal invariant non-topological superconductor (S), the behavior of \JS{$m_2(\phi)$} is such that a quantum phase transition to a state with spontaneously broken time-reversal symmetry at the junction is possible. In this symmmetry-broken state the equilibrium value of the phase takes values $\phi\neq n \pi$ with  $n$ integer. 
 The consequence of such an instability is a peculiar behavior of the Josephson current $J(\phi)$ for small $\phi$, while satisfying the condition
 $J(0)=0$, dictated by time-reversal symmetry.
 %\textcolor{red}{JS: I don't understand this sentence; jump as function of what}.
 %{\color{blue} AS: Sorry, I also do not understand the jump any more. Should think about it. In order to have a jump of $\phi$ we have to control something else. E.g., if we apply a current bias, the phase will jump as in the 1-st order phase transition. The situation was different in our paper with Berg, since there the magnetization has jumped due to the phase bias.} 
 Interestingly, a similar phenomenon takes place in the context of topological Josephson junctions with an embedded magnet. Here, the mass term is generated by the magnetization and a phase transition is predicted to take place when the fluctuations of the magnetic moment are taken into account \cite{reich2023magnetization}.  
One of the goals of the present paper is the analysis of the fluctuations of $\phi$ in the TRITOPS-S junction and the stability of the state with broken time-reversal symmetry.

The second goal is to analyze the effect of vortices in the junction (fluxons). In S-S Josephson junctions, the dynamics of the phase bias is described by the sine-Gordon equation of motion 
and the fluxons lead to the formation of solitons (kink solutions of this equation) where the phase changes by $2\pi$ in the neighborhood of the fluxon \cite{soliton1,soliton2,soliton3,soliton4,soliton5,soliton6,cuevas2014sine}. This profile can propagate along the junction and can collide with other solitons travelling in the opposite direction. We show that the phase dynamics in the TRITOPS-TRITOPS junction can be described in a similar way.
 Instead, the TRITOPS-S junction is governed by a double sine-Gordon equation \cite{campbell1986kink} and a double-kink  solution represents a total change of $2\pi$ in the phase.
In long Josephson junctions of 2D
topological superconductors with broken time-reversal symmetry, the coupling between the solitons and the propagating Majorana edge states has been found to 
localize a Majorana zero mode at the fluxon \cite{gros-stern-pnas}. A similar type of effect is also found to take place in planar Josephson junctions
with magnetic field and spin-orbit coupling hosting a 1D topological superconducting phase \cite{Planarfrac}.

We analyze the impact of single solitons and kink-antikink profiles in 
TRITOPS-S as well as
TRITOPS-TRITOPS junctions. These generate sign changes in the mass terms of the effective Hamiltonian, describing Majorana modes of these junctions (see Fig.~\ref{fig:sketch}). 
 Such a scenario defines an effective Jackiw-Rebbi model \cite{jackreb,goldstone},
which is known to have a localized zero mode. We show that the nature of the zero mode depends on the type of junction. In the TRITOPS-S junction the solitonic solutions are consistent with fractional fluxons that bind 
 Majorana zero modes.
 %Hence, such state is amenable to be manipulated by movingthe fluxon along the junction. 
 The fluxons in the TRITOPS-TRITOPS junction also bind zero modes, which appear in Kramers pairs. 
In the case of kink-antikink profiles in the phase, these localized modes hybridize when they become sufficiently close to one another. Consequently, their energy becomes finite and this effect plays a role in the collisions. In fact,
while ordinary kink and antikink cross one another in a collision\cite{soliton1}, the finite energy of the hybridized modes  generates  an effective repulsion or attraction, which should affect that behavior.

The paper is organized as follows. We present the effective description to analyze the joint dynamics of the phase bias and the Majorana edge modes at the different types of Josephson junctions in Section II. Section III is devoted to analyzing the stability of the mean-field description in the TRITOPS-S junction. The equation of motion for the
phase and the corresponding solitonic solutions in the different junctions are analyzed in Section IV, while Section V is devoted to study the type of electronic bound states which
are generated by these solitonic profiles in the phase. The impact of the electron bound states in the collision of solitons is discussed in Section VI. A summary and conclusions are presented in Section VII. Some technical details are presented in the appendices.

\section{Description of the junction}
We consider the following effective action, which describes the joint dynamics of the phase bias along the junction, $\phi=2\pi \Phi/\Phi_0$ ($\Phi_0=h/2e$), and its coupling to the TRITOPS' edge modes: $S=S_{\phi}+S_{\eta-\phi}$. 
These two components read
\JS{
\begin{equation}
S_{\phi} =  \int_{y,t}\left\{ \frac{\hbar K}{2}\left(\frac{1}{c_{{\rm J}}}\left(\partial_{t}\phi\right)^{2}-c_{{\rm J}}\left(\partial_{y}\phi\right)^{2}\right)-E_{{\rm J}}\left(1-\cos\phi\right)\right\}
\label{eq-action_0a}
\end{equation}
as well as 
\begin{equation}
S_{\eta-\phi}=\frac{1}{2}\int_{y,t} \eta^{\dagger}\left(i\hbar\partial_{t}-{\cal H}\right)\eta,
\label{eq-action_0b}
\end{equation}
with Dirac Hamiltonian 
\begin{equation}
{\cal H}=-i\hbar v\alpha\partial_{y}+m_{1}\left(\phi\right)\beta_{1}+m_{2}\left(\phi\right)\beta_{2}.
\label{eq-action_0c}
\end{equation}
We use the notation $\int_{y,t}\cdots =\int dy\,dt \cdots$. The precise expressions for the matrices $\alpha$ and $\beta_{1,2}$ depend on the nature of the junction and will be given below.
}
%\textcolor{red}{
%\begin{eqnarray}\label{eq-action_0}
%    S_{\phi}&=& \int dy dt\,\left\{\frac{\hbar K}{2}\left(\frac{1}{c_{\rm J}}(\partial_t\phi)^2-c_{\rm J}(\partial_y\phi)^2\right)-E_{\rm J}(1-\cos\phi) \right\}, \\
%    S_{\eta-\phi}&=&\frac{1}{2}\int dy dt\,\eta^\dagger \left\{i\hbar\partial_t+i\hbar {\rm v} \alpha_1\partial_y-m(\phi)\,\alpha_0 - m'(\phi)\,\alpha_0'\right\}\eta.\nonumber
%\end{eqnarray}
%}
$S_{\phi}$ describes the dynamics of the phase, taking into account the effect of the states above the gap only.
% The quantities $\lambda_{\rm J}$ and $\omega_{\rm J}$ are, respectively, the Josephson penetration depth and plasma frequency, while $c_{\rm J}=\lambda_{\rm J} \omega_{\rm J}$ is the Swihart velocity. They depend on the amplitude of the Josephson current $J(\phi)=J_0 \sin(\phi)$, $J_0=2e E_{\rm J}/\hbar$, the capacity $C$ of the junction and the penetration $d=d_0+\lambda_1+\lambda_2$, where $d_0$ is the distance between the two superconductors, while $\lambda_1, \lambda_2$ are the London penetration lengths. Explicitly,  $\omega_{\rm J}=\sqrt{2 e J_0/\hbar C}$ and $\lambda_{J} =\sqrt{\hbar c^2/(8 \pi e d J_0)}$.
$E_{\rm J}$ is the Josephson energy density per length, which is related to the amplitude of the Josephson current density (per length) $J(\phi)=J_0\sin(\phi)$ via $J_0 = 2eE_{\rm J}/\hbar$. 

The gradient contributions to 
$S_{\phi}$ originate in the energies of the electric and magnetic fields. The 
action assumed here corresponds to a 3D situation, i.e. a junction of sufficiently large height $h_z \gg \lambda_{1,2}$, where $\lambda_{1,2}$ are the respective London penetration depths. In this regime the junction forms a wave guide with most of the magnetic and electric energies bound to its volume. The phase rigidity $K$ and the effective light velocity (Swihart velocity) $c_{\rm J}$ are then determined by the geometry of the junction and given by $c_{\rm J}^2 = c^2 d_0/d$, $K=\hbar c h_z/(16 \pi e^2 \sqrt{d d_0})$. Here $d_0$ is the width of the insulating barrier separating the two superconductors and $d=d_0+\lambda_{1}+\lambda_{2}$ \cite{Hermon1994}. 
If $h_z$ becomes too small, $h_z \ll \lambda_{1,2}$, stray magnetic fields extending outside of the junction need to be considered, leading to a non-local effective action \cite{Abdumalikov_2009}. 

Although the effective models for the topological superconductors considered in the present work are 2D, the mechanisms to generate such phases are typically by proximity to 3D superconductors\cite{fu2010odd,zhang2013time,keselman2013inducing,haim2016interaction,
 reeg2017diii,santos2010superconductivity,klinovaja2014kramers,mellars2016signatures,parhizgar2017highly,casas2019proximity,iron,haim2019time}. We assume these 3D superconductors to provide the necessary suppression of the stray magnetic fields through their height $h_z$, acting as a wave guide and allowing us to consider a purely local theory. Note that in the usual 3D Josephson junctions the Josephson energy density $E_{\rm J}$ per length is also proportional to $h_z$ \cite{Hermon1994, Tinkham}. In that case, one introduces the Josephson energy density per area so that the height $h_z$ becomes a prefactor of the whole action $S_{\phi}$, which can effectively be reformulated such that $K$ is the prefactor of $S_{\phi}$.
In contrast, here the Josephson tunneling takes place only between the 2D superconducting layers, thus our Josephson energy density per length, $E_{\rm J}$, is completely unrelated to $K$. 

%$K$ is the dimensionless stiffness of the phase modes. In a classical junction where the superconductors and the junction are three-dimensional of finite height $W$, this parameter is given by $K^2=\hbar c W/(16 \pi e^2 \sqrt{d d_0})$ \cite{thesis,coleman1975quantum}. In such a classical junction, the mean-field description for $\phi$ is expected to be reliable. This is associated to a large value of $W$, hence, large $K$. The opposite limit, when $K$ is small, represents a scenario with large quantum fluctuations of the phase and this regime is approached as the superconductor becomes ideally 2D.

%The effective models for the TRITOPS superconductors considered in the present work are 2D. However, the mechanisms to generate this phase are typically by proximity to 3D superconductors. Hence, we consider $1/K$ in the forthcoming analysis as a phenomenological parameter that determines the ''quantumness'' of the phase dynamics where small (large) values of $K$ describe strong (weak) quantum fluctuations of the phase.

$S_{\eta-\phi}$ describes the coupling between the Majorana edge states and the phase. The details depend on the type of junction, which is characterized
by the  mass terms $m_{1,2}(\phi)$ and the
number of Kramers pairs of counterpropagating
Majorana modes involved. The latter defines the structure of the spinor $\eta$ (see Ref. \onlinecite{us}). $v$ is the velocity of the Majorana modes.

%\LA{*****PLEASE CHECK THIS***** }

\JS{In the action introduced in Eqs.~(\ref{eq-action_0a})-(\ref{eq-action_0c}) we consider two mass terms. The first one is denoted by $m_1(\phi) \beta_1$ and corresponds to a coupling between the edge modes on both sides of the junction. Its leading contribution is of first order in the tunnelling element $t_{\rm J} $.
%This term is generated by $n$-th order tunneling processes modulated by $t_{\rm J} e^{\pm i \phi/2}$. Taking into account the symmetries of the junction, this term can have the form
%\begin{eqnarray}
%    & & \cos(n\phi/2) \tau^x \sigma^0 \;\;\; {\rm or} \;\;\; \cos(n\phi/2) \tau^y \sigma^j,\;j=x,y,z,\nonumber \\
%    & & \sin(n\phi/2) \tau^y \sigma^0 \;\;\; {\rm or} \;\;\; \sin(n\phi/2) \tau^x \sigma^j,\;j=x,y,z.
%\end{eqnarray}
%Notice that the terms of the first line  are modulated by an even function of $\phi$, the matrices $\alpha_0$ are TRI. Hence, all these terms  are TRI $\forall \phi$. 
%This term contains the leading order $n=1$ and the different combinations of Pauli matrices depend on the type of hopping (spin-preserving or with  spin flip). 
%Instead, the terms of the second line are modulated by an odd function of $\phi$. Hence, they contain matrices  $\alpha_0$ transforming to  $-\alpha_0$ under TR. 
The second mass term in Eq.~(\ref{eq-action_0c}), denoted by $m_2(\phi) \beta_2$, is due to  the virtual coupling of edge modes on one side  of the junction to states above the gap on the other side. Hence, it effectively couples the edge modes within each superconductor and is of order  $t_{\rm J}^2/\Delta_{\rm eff} $ with $\Delta_{\rm eff} $ being the magnitude of the superconducting gap. Obviously, for the TRITOPS-S junction $m_2$ is the only possible mass term.}
%Here, the only allowed terms are of the form 
%\begin{equation}
%    \sin(n\phi/2) \tau^z \sigma^j, \;\;\; \sigma^j,\;j=x,y.
%\end{equation}
%We exclude terms with $\cos(n\phi/2)$ because they break TR symmetry (notice that we need $\sigma^x$ or $\sigma^y$). We also exclude terms with $\sin(n\phi/2) \tau^0 \sigma^j$ because they are not invariant under the change
%$\phi \leftrightarrow -\phi$ and ${\rm S}_1 \leftrightarrow {\rm S}_2$.

%\LA{*******END*******}

We now summarize the description  on the basis of low-energy Hamiltonians describing a spin-preserving
tunneling  ($t_{\rm J}$) at the Josephson junction, following the derivation presented in Ref.~\onlinecite{us}.  
For the
TRITOPS-S junction there is a single pair of counter propagating Majorana modes (see Fig.~\ref{fig:sketch}). Consequently, $\eta$ is a two-component spinor 
and we have
\JS{
\begin{gather}\label{intro_TRITOPS-S}
%\eta(y)&=&\left(\eta_{\uparrow}(y), \eta_{\downarrow}(y) \right)^T, \\
%& & {\rm TRITOPS-S:}\nonumber\\
%& & \beta_2=\sigma^y,\; \alpha=\sigma^z,\nonumber\\
 \alpha= \sigma^z, \quad
  \beta_2=\sigma^y,\\
 m_1(\phi)=0,\quad
 m_2(\phi)=m_2^{(0)} \sin(\phi).\label{eq:massTS}\\[.7em]
 {\rm (TRITOPS-S)} \nonumber
\end{gather}
}
The Pauli matrices $\sigma$ act on the two members of the Majorana Kramers pair. The edge modes of the TRITOPS are coupled to the supragap states of the non-topological superconductor S. As mentioned before,
the non-vanishing mass term  $m_2^{(0)}\propto t_{\rm J}^2/\Delta_{\rm eff}$
is originated in a second-order process in  the tunneling  amplitude. 
%$t_{\rm J}$, being  $\Delta_{\rm eff}$ the effective pairing amplitude in the topological phase.

In distinction, for the TRITOPS-TRITOPS junction (also sketched in Fig.~\ref{fig:sketch}), $\eta$ is a spinor with four components corresponding to the two Kramers pairs associated to the two TRITOPS (labeled by ${\rm S}_1,{\rm S}_2$),
and we have
\JS{
\begin{gather}\label{intro_TRITOPS-TRITOPS}
%& & {\rm TRITOPS-TRITOPS:}\nonumber\\
%\eta(y)&=&\left(\eta_{1,\uparrow}(y), \eta_{1,\downarrow}(y),\eta_{2,\uparrow}(y),\eta_{2,\downarrow}(y)\right)^T, \\
\alpha=\tilde{\sigma}^z  \sigma^z, \quad \beta_1=\tilde{\sigma}^y  \sigma^z,\quad \beta_2=\tilde{\sigma}^0\sigma^y,\\
m_1(\phi)=m_1^{(0)} \cos(\phi/2),\quad m_2(\phi)=m_2^{(0)}\sin(\phi). \label{eq:massTT}\\[.7em]
{\rm (TRITOPS-TRITOPS)} \nonumber
\end{gather}}
$\tilde{\sigma}^j$ are Pauli matrices acting in the ${\rm S}_1,{\rm S}_2$ subspace while $\sigma^j$
act on
the two partners of the Kramers pair. Here, the mass term modulated by $m_1^{(0)} \propto t_{\rm J}$ is generated by the hybridization of the Majorana edge states at both sides of the junction. 
This is the leading order in the tunnelling $t_{\rm J}$ and it is exact when the superconducting gap  is large enough $\Delta_{\rm eff} \gg t_{\rm J}$ to prevent the hybridization of the edge modes on one side of the
junction with the supragap states of the other side. The second mass term $m_2(\phi)$ is a second-order perturbation with $m_2^{(0)}\propto t_{\rm J}^2/\Delta_{\rm eff}$
and has to be taken into account when the 
edge states on one of the TRITOPSs hybridize, not only with the edge states,  but also with the supragap states of the other TRITOPS. 

\JS{To identify the $4\times4$ matrices $\alpha$ and $\beta_{1,2}$ of Eq.~\eqref{intro_TRITOPS-TRITOPS} that determine the Dirac Hamiltonian ${\cal H}$ for the TRITOPS-TRITOPS
junction, we analyze the behavior under time reversal, charge conjugation,
and under the exchange of the two identical TRITOPSs. Time reversal
corresponds to simultaneously transforming $\phi\rightarrow-\phi$
and ${\cal T}=i\tilde{\sigma}^{0}\sigma^{y}{\cal K}$ in the fermion
sector. The charge conjugation
operator $\mathcal{C}=U_{C}\mathcal{K}$ acts on the BdG Hamiltonian
as $U_{C}{\cal H}U_{C}^{-1}=-{\cal H}^{\ast}$. We use a basis with real
field operators  $\eta^\dagger = \eta$. 
In this basis  $U_{C}=1$ and thus  ${\cal H}=-{\cal H}^{\ast}$.
Finally, the  exchange symmetry of the two superconductors S$_1$ and S$_2$
corresponds to simultaneously transforming $\phi\rightarrow-\phi$
and $U_X=\tilde{\sigma}^{x}\sigma^{x}$ in the fermion
sector. The conditions of time-reversal, charge-conjugation, and exchange symmetry
uniquely determine the matrices for the TRITOPS-TRITOPS
junction. In Appendix \ref{appendix_symmetries} we give further details on the derivation of these symmetries.}
 The underlying assumptions are $p_{\pm}$ type of pairing in both superconductors and spin-preserving tunneling at the junction. 

\JS{The transformation $\phi\rightarrow-\phi$ is necessary if the phase
$\phi$ is a genuine dynamic variable. However, if one considers  the properties of the fermionic sector for a
fixed configuration $\phi\left(y\right)$,  the mass term $m_{2}\left(\phi\right)$
breaks both, the time-reversal   and the exchange symmetry. This will be important when we analyze the fermionic spectrum near given soliton solutions of the phase. }

%Notice that this term is invariant under the symmetry ${\cal S}$ that corresponds to change $\uparrow \leftrightarrow \downarrow$ with a simultaneous exchange of  S$_1$ and S$_2$, which is a symmetry of the junction. Similarly, it is also invariant under a simultaneous change $\phi \rightarrow -\phi$ and
%an exchange between
%S$_1$ and S$_2$. The mass term with $m_0$ preserves time-reversal symmetry generated by the operator ${\cal T}={\cal K}i\tilde{\sigma}^0 \sigma^y$, where
%${\cal K}$ denotes the complex conjugation operation. Notice that ${\cal T}^{-1} \sigma^z {\cal T} = - \sigma^z$ and ${\cal T}^{-1} \tilde{\sigma}^y {\cal T}=-\tilde{\sigma}^y$.
%The kinetic term and the mass term $m_0$ also preserve another time-reversal symmetry generated by the operator $\tilde{\cal T}={\cal K}i\tilde{\sigma}^y \sigma^0$. }%In addition,${\cal T} \times \tilde{\cal T}$ is also a symmetry.
%\textcolor{red}{
%Other higher order terms in the tunneling generating a massive term that breaks TR symmetry while preserves the symmetry ${\cal S}$ introduced before are
%$m'_j \sin (n\phi/2) \alpha'_j, \; j=1,\ldots,3$, being $n$ the order of the tunneling process and
%\begin{eqnarray}
%    & & \alpha'_1=\tilde{\sigma}^z\sigma^x, \;\;\; \alpha'_2=\tilde{\sigma}^z\sigma^y, \;\;\;\alpha'_3=\tilde{\sigma}^y\sigma^0
%\end{eqnarray}
%}

\JS{The distinct dependency of the edge-mode mass $m_{1,2}(\phi)$ for the two junctions, given in Eqs.~\eqref{eq:massTS} and \eqref{eq:massTT}, gives rise to qualitatively distinct behavior. For the TRITOPS-TRITOPS junction, the net mass at $\phi=0$ is finite while it vanishes for the TRITOPS-S junction. This is 
 consistent with the bulk-boundary correspondence; see Ref.~\onlinecite{us} for a discussion. As we will see, a fermion zero mode will only be associated with a phase slip soliton if the fermion is massive at constant phase. For the TRITOPS-S, this requires spontaneously breaking time-reversal symmetry with an equilibrium phase $\tilde{\phi} \neq n \pi$. In this case, the protecting symmetry of the  massless edge modes is spontaneously broken for the TRITOPS-S junction and the  bulk-boundary correspondence does not apply.}
 %\GRR{****In the last version of our paper we have stated that bulk-boundary correspondence was broken****}  A such that $m(\tilde{\phi})\propto \sin \tilde{\phi} \neq 0$.
We will see below that the stability of the time-reversal symmetry broken state depends on the value of the rigidity $K$. The symmetry is broken for large values of $K$ and hence large values of $h_z$, while at small $K$ quantum fluctuations are important and restore the symmetry. The ordered state is characterized by an Ising variable that describes the two states $\pm \tilde{\phi}\, {\rm mod}(2\pi)$.  The quantum phase transition is then expected to be in the tricritical Ising universality class\cite{grover_emergent_2014,rahmani_emergent_2015}, the natural transition from an Ising ordered phase to a massless, critical phase that corresponds to the critical point in the usual Ising model. It is described by the tricritical Ising conformal field theory with central charge $7/10$.
In our analysis we are more interested in the quantitative location and parameter dependence of this transition and in the properties of phase slips on the ordered side of the transition. The ordered state can be described in a controlled fashion in the limit of large rigidity $K$. To estimate the phase boundary we will then go beyond the mean field limit, valid at $K\rightarrow \infty$, and include fluctuation effects; the transition is then estimated from the parameter regime where these fluctuations start to dominate the equation of state.

\section{Large phase rigidity and role of fluctuations}
We now consider a semi-classical approach, where the fermionic modes are treated in a full quantum mechanical framework, 
while $\phi(y,t)$ is regarded as a classical field defined by its mean value with $K\rightarrow\infty$ in Eq.~(\ref{eq-action_0a}).
Under this assumption, we express $\eta(y)=\sum_{k}  e^{i k y}\eta_k $ and we consider the Hamiltonian \JS{in Eq.~\eqref{eq-action_0c}
%\begin{equation}\label{heff-class}
%H_{\eta-\phi} = \sum_{k\geq 0} \; \eta^{\dagger}(k)\left[ m(\phi) \alpha_0 - {\rm v} \alpha_1 k \right]\eta(k),
%\end{equation}
where the masses $m_{1,2}(\phi)$ are functions of the classical field $\phi$, specified in Eqs.~\eqref{eq:massTS} and \eqref{eq:massTT}. }The total energy density after integrating out all the fermionic modes of the junction in $\langle H_{\eta-\phi} \rangle$ and adding the contribution of the supragap states reads
\begin{equation}\label{ephi}
E(\phi) = E_{\rm J}\left[1-\cos(\phi)\right]-\frac{m^2(\phi)}{8 \pi\hbar v} \left[1+
 \log\left( \frac{4\Lambda^2}{m^2(\phi)}\right) \right],
\end{equation}
\JS{where $m^2(\phi)=m^2_1(\phi)+m^2_2(\phi)$ is the net mass and $\Lambda$ a high-energy cutoff of the order of the effective superconducting gap $\Delta_{\rm eff}$.}

The contribution of the states above the superconducting gap, i.e. all states except for the edge modes, is given by the first term of Eq.~(\ref{ephi}) which  has a minimum at $\phi=0$.
As discussed above, the effect of the edge modes given by the second term does not affect this minimum in the TRITOPS-TRITOPS 
junction since $m(\phi) \propto m_0$ at small $\phi$, while  a competing minimum may 
develop for the TRITOPS-S junction~\cite{us}. In what follows we analyze the stability of this time-reversal symmetry breaking solution in the TRITOPS-S junction against the effect of the phase fluctuations introduced by a finite $K$ in Eq.~(\ref{eq-action_0a}). 

Under the assumption $\phi\ll 1$, one finds the minima of the ground state energy-density (\ref{ephi}) to be 
\begin{equation}\label{phieq}
\phi_\text{\rm eq}^{(1/2)} = \pm\frac{2\Lambda}{m_0}\exp\left(-\frac{\pi\hbar v E_{\rm J}}{ m_0^2}\right)\neq 0,
\end{equation}
hinting, as discussed, at a spontaneous symmetry breaking. However, one has to keep in mind the possibility of quantum fluctuations of $\phi$ destroying the phase with $\phi_{\rm eq}\neq 0$. This is exemplified by expanding the action in Eq.~\eqref{eq-action_0a} up to second order in $\phi$, which yields the Gross-Neveu-Yukawa (GNY) model in 1+1 dimensions, a bosonized version of the Gross-Neveu model \cite{gross_dynamical_1974}, $S\approx S^{(0)}_{\phi}+S_{\eta-\phi}$ with
 \begin{equation}
S_{\phi}^{\left(0\right)}=\int_{y,t}\text{\ensuremath{\left\{ \frac{\hbar K}{2}\left(\frac{1}{c_{{\rm J}}}\left(\partial_{t}\phi\right)^{2}-c_{{\rm J}}\left(\partial_{y}\phi\right)^{2}\right)-\frac{E_{{\rm J}}}{2}\phi^{2}\right\} }}.\label{eq-GNY-action}
 \end{equation}
%\begin{eqnarray}\label{eq-GNY-action}
 %   S=\int dy\,dt\,\left\{\frac{1}{2}\eta^\dagger\left(i\hbar\partial_t+i\hbar {\rm v} \alpha_1\partial_y-m_0\phi\,\alpha_0\right)\eta\right. \nonumber \\
  %  \left.+\frac{\hbar K}{2}\left(\frac{1}{c_{\rm J}}(\partial_t\phi)^2-c_{\rm J}(\partial_y\phi)^2\right)-\frac{E_{\rm J}}{2}\phi^2\right\}.
%\end{eqnarray}
In the limit $K\rightarrow 0$, integrating out $\phi$ results in a local 4-Majorana interaction which is RG irrelevant and thus no symmetry breaking takes place: in this limit, the mean-field approach is invalidated by fluctuations. Since the stiffness $K$ suppresses the fluctuations (and renders the effective 4-Majorana interaction non-local), we expect a quantum phase transition at some positive value of $K$, beyond which the non-zero expectation value of $\phi$ is stabilized. 

While in the GNY model the question regarding a symmetry broken phase poses a strong-coupling problem, formally not accessible via perturbative methods, it has been shown in Ref.~\onlinecite{reich2023magnetization} that one can derive an estimate for the parameter regime in which symmetry breaking is allowed by analyzing the corrections to the equation of state $dE(\phi)/d(\phi^2) = 0$ 
due to Gaussian fluctuations. 

Following the reasoning of Ref.~\onlinecite{reich2023magnetization}, we expand the GNY action up to second order in the fluctuations $\delta\phi$ around some assumed minimum $\braket{\phi}=\tilde{\phi}$ of the ground state energy, $\phi(x,t)=\tilde{\phi}+\delta\phi(x,t)$. After integrating out $\delta \phi$, we derive the equation of state for $\tilde{\phi}$ and examine whether it allows for solutions $\tilde\phi\neq 0$, i.e.\ whether the mean-field solution is consistent and only weakly affected by fluctuations. If this is the case, we conclude that fluctuations are small and the treatment in the limit of large rigidity  therefore justified. If not,  fluctuations are large, suppress  long-range order, and the symmetry remains unbroken. 

\begin{figure}
    \centering
    \includegraphics[width=\linewidth]{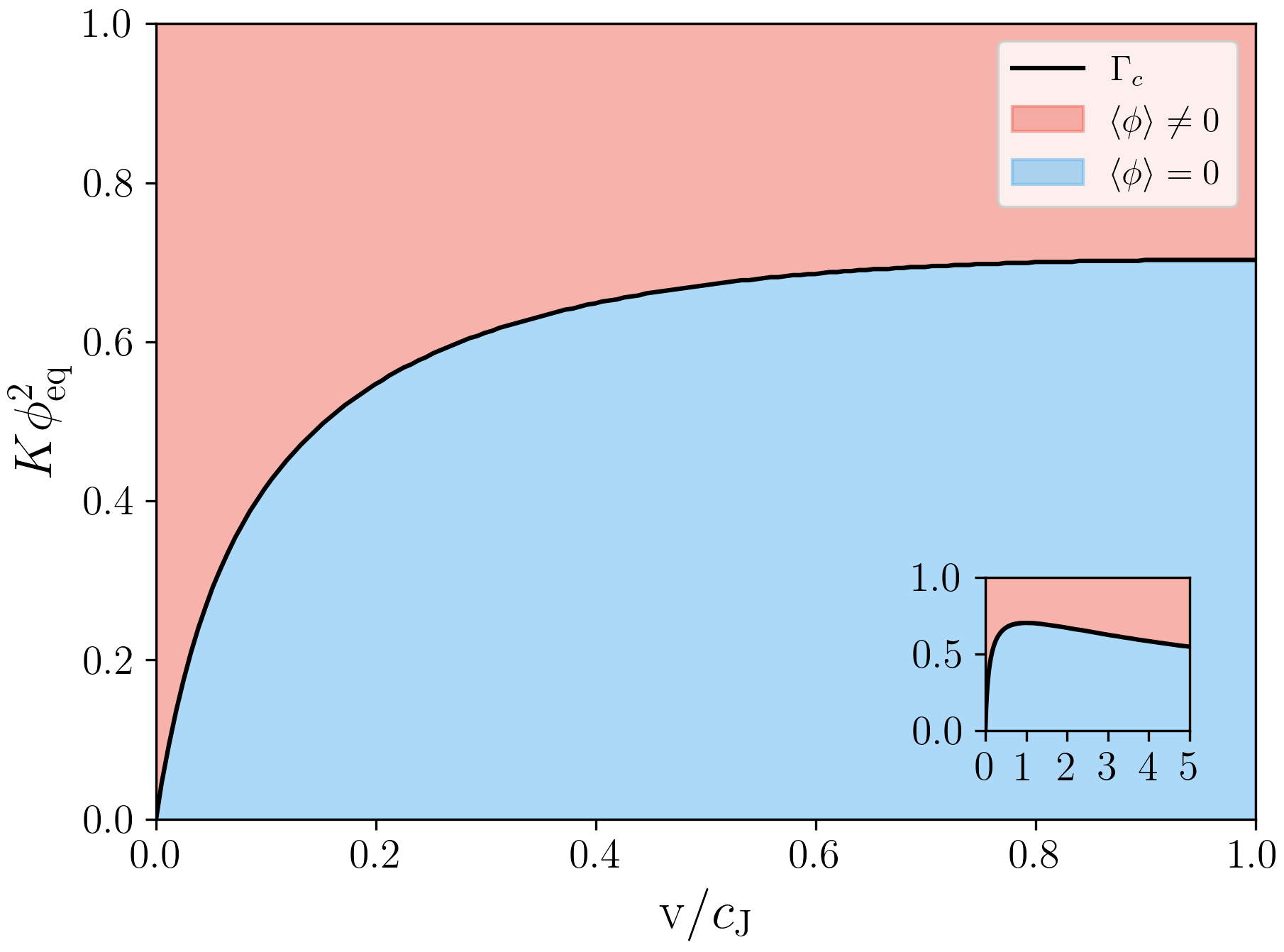}
    \caption{Phase diagram of the TRITOPS-S junction with the boundary $\Gamma_c$ between symmetry broken (red) and unbroken (blue) phase determined from Eq.~(\ref{gap-eq_fluct}) for ${v}/c_{\rm J}\in [0,1]$ (inset: ${v}/c_{\rm J}\in [0,5]$).}
    \label{fig:phaseboundary_fromfluct}
\end{figure}

The fluctuation-corrected equation of state determining the gap in the Majorana spectrum, the derivation of which is sketched in Appendix \ref{appendix_gapeq}, reads 
\begin{equation}\label{gap-eq_fluct}
    \frac{dE}{d(\tilde\phi^2)}=0\quad\Leftrightarrow\quad \log\left|\frac{\tilde\phi}{\phi_{\rm eq}}\right|+\chi_{\tilde{\phi}}=0,
\end{equation}
with
\begin{align}
    \chi_{\tilde{\phi}} = \frac{4}{\pi}\int_0^{\pi/2}&d\theta\int_0^\infty dr\,\frac{\rm{Arsinh}(r)}{\sqrt{1+r^2}}\left[8\pi K\tilde\phi^2r^2\left(\frac{v}{c_{\rm J}}\sin^2\theta\right.\right.\\
    +&\left.\left.\frac{c_{\rm J}}{v}\cos^2\theta\right)+\log\left|\frac{\tilde\phi}{\phi_{\rm eq}}\right|+\frac{\sqrt{1+r^2}}{r}\rm{Arsinh}(r)\right]^{-1}. \nonumber
\end{align}
Here, $(r,\theta)$ are polar coordinates in the $\left(\frac{\hbar\omega}{2m_0\tilde\phi},\frac{{\hbar v} q}{2m_0\tilde\phi}\right)$-plane. We calculate the integral numerically for different values of ${v}/c_{\rm J}$ and find the curve of critical values $\Gamma_c({v}/c_{\rm J})$ depicted in Fig.~\ref{fig:phaseboundary_fromfluct}, representing the minimum value $K\phi_{eq}^2$ needs to take on in order for solutions to Eq.~(\ref{gap-eq_fluct}) to exist and thus the symmetry breaking to be stable against Gaussian fluctuations.

The relevant experimental scenario corresponds to small values of ${v}/c_{\rm J} \ll 1 $ ($\simeq 10^{-4}-10^{-3}$), in which case a small value of $K \phi_{\rm eq}^2$ is enough to guarantee the stability of the mean field solution. The precise value of $K$ should anyway depend on the geometric details of the junction. 

Within the Gaussian fluctuation  framework, the  quantum phase transition is  of first order. However, there is no reason to expect that  the employed approximation properly describes the critical fluctuations at this transition correctly. As stated above, the quantum phase transition is expected to be in the tricritical Ising universality class, where all exponents are exactly known. Our approach is however very useful in getting an estimate of the phase boundary as function of the system parameters.

\section{Equation of motion for the phase bias and solitonic solutions}
We now focus on the regime of large $K$ of Fig.~\ref{fig:phaseboundary_fromfluct}, where the fluctuations of the phase bias are not dominant so that we can regard $\phi(y,t)$ as a classical variable leading to a current-phase relation (CPR) $J(\phi)$.
We assume that the length of the junction is significantly larger than the Josephson penetration depth of the two superconductors, $\lambda_1, \lambda_2$. Hence, the phase may have a spatial dependence along the junction which is represented by $\phi(y,t)$, whose dynamics is described by 
%a Hamiltonian $H^{\phi}$, which depends of the capacitance $C$ of the junction and the CPR. The corresponding 
the following equation of motion 
%\begin{equation}\label{eq-mov}
%\frac{\hbar  {\color{red}h_z}}{8 \pi e d_0}\partial^2_t {\phi}- \frac{\hbar c^2 {\color{red}h_z}}{8 \pi e d}  \partial_y^2 {\phi} +J(\phi) =-\frac{\hbar G_N}{2 e } \partial_t \phi  + J_{\rm ext},
% - \beta \partial_t \partial^2_y \phi 
%\end{equation}

\begin{equation}\label{eq-mov}
\frac{\hbar}{2e} C \partial^2_t {\phi}- 
\frac{\hbar}{2e} \frac{1}{L}  \partial_y^2 {\phi} +J(\phi) =-\frac{\hbar G_N}{2 e } \partial_t \phi  + J_{\rm ext}.
% - \beta \partial_t \partial^2_y \phi 
\end{equation}
%\textcolor{olive}{Which of these two possibilities do we prefer?}
Here, $C=h_z/(4\pi d_0)$ and $L=4\pi d/(h_z c)^2$ are the capacitance per length and the inductance per length of the wave guide, respectively. 
The right-hand side of the equation describes the 
dissipation due to the normal conductance (per length) $G_N$
and the driving with an external current density $J_{\rm ext}$.
% while $\beta$ describes the surface resistance of the superconductors due to the electronic propagation along the junction.

The homogeneous equation  corresponding to Eq.~(\ref{eq-mov}) with the right-hand side equal to zero defines the
sine-Gordon equation \cite{soliton1} for  $J(\phi) = J_0 \sin(\phi)$.
%, with $J_0=2e E_{\rm J}/\hbar$. 
This
is the equation of motion derived from the action $S_{\phi}$ of Eq.~(\ref{eq-action_0a}). 
In fact, one can introduce new length and time scales, $\lambda_{\rm J}$ and $\omega_{\rm J}^{-1}$, and redefine $y/\lambda_{\rm J} \rightarrow y$, $t \omega_{\rm J} \rightarrow t$, such
%, with $\omega_{\rm J}=\sqrt{2 e J_0/\hbar C}$, $\lambda_{J} =\sqrt{\hbar c^2/(8 \pi e d J_0)}$ and $c_J = \sqrt{c^2/(4 \pi C d)}$. 
that the homogeneous equation simply reads
\begin{equation}\label{sine-g}
\phi_{t t } - \phi_{yy} + \sin \phi =0.
\end{equation}

It admits soliton -- or kink -- solutions associated to fluxons trapped in the junction \cite{soliton1}. These solutions have the following structure,
\begin{equation}\label{kink}
\phi(y,t)= 4 \arctan \left(e^{\frac{y -u t }{\sqrt{1-u^2}}} \right),
\end{equation}
being $u$ the velocity at which the soliton propagates. Its precise value is determined by considering the full
Eq.~(\ref{eq-mov}), including the right-hand side. 

We now focus on the static limit of Eq.~(\ref{eq-mov}) in the homogeneous case ($u=0$) and we defer the discussion of the dynamical effects to Section \ref{dyn}.
Expressing the CPR as $J(\phi)=E_{\rm J} \partial_\phi \epsilon(\phi)$,
this equation can be written  as,
\begin{equation}\label{sol-prof}
\frac{1}{2} \frac{d}{dy} \left(\frac{d \phi}{dy} \right)^2= \frac{d \epsilon(\phi)}{dy}.
\end{equation}
Taking into account the boundary conditions $d\phi(y)/dy=0,  \;y \rightarrow -\infty$ and  $\phi(y)=\phi_{\rm eq}, \;y \rightarrow -\infty$, being $\phi_{\rm eq}$ the phase corresponding to the
minimum of $\epsilon(\phi)$, Eq.~(\ref{sol-prof}) can be also  expressed as follows
\begin{equation} \label{eq-dif}
\frac{d \phi}{d y} =\pm \sqrt{2 \delta \epsilon(\phi)}, \;\;\;\;\; \delta \epsilon(\phi)= \epsilon(\phi)-\epsilon(\phi_{\rm eq}).
\end{equation}
It is easy to verify that for $\epsilon(\phi)= 1 - \cos(\phi)$, we recover the stationary sine-Gordon equation
with the solution given by Eq.~(\ref{kink}).

%\section{Sine-Gordon equations for the different junctions}

Our aim is to  solve Eq.~(\ref{sol-prof}) for  different configurations of
Josephson junctions containing one or two TRITOPSs.

\subsection{TRITOPS-S junction}
The remarkable feature of this junction is the fact that the phase corresponding to the minimum energy, $\phi_{\rm eq}$, has a finite value 
 $0 < \phi_{\rm eq}\leq \pi/2$, meaning a spontaneous symmetry breaking for arbitrarily small values of the phase bias.
 
\begin{figure}[t]
\centering
\includegraphics[width=\linewidth, trim={0.5cm 0.7cm 0cm 0.5cm},clip]{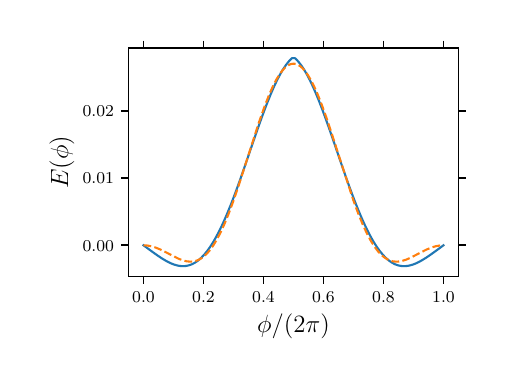}
\caption{Plot in continous lines: Energy vs.\ phase for a junction ${\rm S}_1-{\rm S}_2 \equiv {\rm TRITOPS}-{\rm S}$ calculated in the lattice model defined in Appendix \ref{aplat} with $N_x=N_y=300$ sites and periodic boundary conditions along the junction.
Parameters are: the tunneling hopping at the junction $t_J=w/5$, the amplitude of the pairing potentials of the two superconductors, $\Delta_s^{(2)}=\Delta_p^{(1)}=4\Delta_s^{(1)}=w/5$, and the chemical potential $\mu=-2w$, being $w$ the hopping parameter of the square lattice. Plot in dashed lines: Eq.~(\ref{ephi-sin2}) with $E_0=0.006$ and $E_{\rm J}=0.014$.}\label{energy}
\end{figure}

In order to address the related properties of the junction, we focus  on specific parameters in Eq.~(\ref{ephi}), which lead to an integrable double sine-Gordon equation \cite{campbell1986kink} for the phase bias. This corresponds to assuming 
 $\Lambda \gg m_0$ in Eq.~(\ref{ephi}), so that we can substitute the term with the $\log$ function by a constant. The result is
 \begin{equation}\label{ephi-sin2}
 E(\phi)\simeq 
 %E_{\rm J}(1-\cos(\phi))-2 E_0 \sin^2(\phi) = 
 E_{\rm J}(1-\cos(\phi))-E_0 \left[1-\cos(2\phi)\right].
 \end{equation}
%$E(\phi)\simeq E_{\rm J}(1-\cos(\phi))-2 E_0 \sin^2(\phi) = E_{\rm J}(1-\cos(\phi))-E_0 \left[1-\cos(2\phi)\right]$.
 This energy has two minima at 
 \begin{equation}\label{mini}
 \phi_{\rm eq}^{(1)}=\phi_0, \quad \phi_{\rm eq}^{(2)}=2 \pi - \phi_0,\quad
 \phi_{0}=\arccos\left(\frac{E_{\rm J}}{4E_0}\right).
 \end{equation}
 We see from the definition of $\phi_0$ that the two solutions, $\phi_{\rm eq}^{(j)},\;j=1,2$,
exist provided that $4E_0>E_{\rm J}$. Otherwise, we get the single (usual) solution $\phi_{\rm eq}=0,\; \mbox{mod}[2\pi]$.
We benchmark this description against the  numerical solution of a lattice model, where
$E(\phi)$ is calculated in a junction between two BCS Hamiltonians defined in a square lattice connected in a Josephson arrangement following Ref.~\onlinecite{us} (see details in Appendix \ref{aplat}).
One of the Hamiltonians (corresponding to the superconductor S$_1$) is a TRITOPS  with p-wave and s-wave pairing,
respectively $\Delta_p^{(1)},\;\Delta_s^{(1)}$,
and chemical potential $\mu$ within the topological phase, where this system has propagating Majorana modes at the edges. The motivation for
considering these two components of the pairing is the fact that many proposals for achieving the TRITOPS phase in real materials rely on
a combination of s-wave superconductivity with spin-orbit coupling (see for instance Refs.~\onlinecite{haim2019time,zhang2013time,casas2019proximity,iron,us}), which effectively induces p-wave and s-wave pairing. 
The model for the non-topological Hamiltonian has only s-wave pairing $\Delta_s^{(2)}$.
In order to minimize finite-size effects, periodic boundary conditions are implemented along the direction $y$, parallel to the junction. The dimension of the lattice is  $N_x=N_y=300$ sites for
each of the superconductors. The ground state energy shown in Fig.~\ref{energy} is calculated by
 adding all the negative single-particle energies of the Bogoliubov-de Gennes Hamiltonian. We clearly see two minima corresponding to $\phi_0 \neq 0$. The precise value depends on the model parameters, mainly on $\Delta_p^{(1)},\;\Delta_s^{(1)}$.
 We compare 
these numerical results with those calculated from Eq.~(\ref{ephi-sin2}) with parameters $E_0$ and $E_{\rm J}$ optimized to fit the numerical data.  A good agreement is found in the results shown in the figure. This supports  Eq.~(\ref{ephi-sin2}) as a representative limit to analyze the phase dynamics of the TRITOPS-S junction.

\begin{figure}[t]
\centering
\includegraphics[width=\linewidth, trim={0cm 0.4cm 0cm 0.4cm},clip]{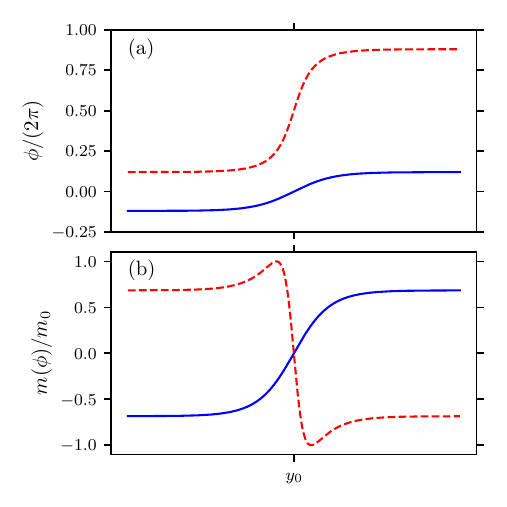}
\caption{Profile for a soliton of the phase (a) and the mass \JS{$m_2(\phi)$}  (b) for the two types of solution $\phi_1$ and $\phi_2$ of the TRITOPS-S junction represented by continuous and dashed lines with $\phi_0=0.12\times 2\pi$.}\label{mass-fig-T-S}
\end{figure}

%\subsubsection{Solitonic solutions}
We now turn to discuss the possibility of having solitonic solutions to Eq.~(\ref{eq-dif}) in this peculiar junction. 
We focus on the situation where we get the two minima given by Eq.~(\ref{mini}) and we express $\delta\epsilon(\phi)=\left[E(\phi)-E(\phi_{0})\right]/E_{\rm J}$ as
%and considering $\delta\epsilon=\frac{E(\phi)-E(\phi_{0})}{E_{\rm J}}=-cos(\phi)+cos(\phi_{0})+\frac{E_0}{E_{\rm J}}\left(cos(2\phi)-cos(2\phi_{0})\right)$. \GRR{Since $\phi_{0}=arccos(\frac{E_{\rm J}}{4E_0})$, for $4E_0>E_{\rm J}$, we obtain 
\begin{eqnarray}\label{TS}
\delta\epsilon (\phi)
%&=&- \cos(\phi)+\frac{E_0}{E_{\rm J}} \cos(2\phi)+\frac{E_{\rm J}}{8E_0}+\frac{E_0}{E_{\rm J}}\nonumber \\
& = & -\cos(\phi)+\frac{1}{2\cos(\phi_0)}\left[\cos^2(\phi)+\cos^2(\phi_0)\right].
\end{eqnarray}
This defines an integrable double  sine-Gordon equation \cite{campbell1986kink}.
 It has  two solitonic solutions, which correspond to solving
\begin{equation}
s\int_{\overline{\phi}_j}^{\phi_j} d \varphi \frac{1}{\sqrt{2 \delta \epsilon(\varphi)}}= y-y_0, \;\;\;
\end{equation}
with $s=\pm$, $\phi_1(y_0)=\overline{\phi}_1=0$ and $\phi_2(y_0)=\overline{\phi}_2=\pi$. They read
\begin{eqnarray}\label{eq:non_trivial_kinks}
    &\phi_1(y) = 2\arctan \left[ \tan\left(\frac{\phi_{0}}{2}\right)\tanh\left( \frac{|\sin \phi_{0}|}{2\sqrt{\cos(\phi_0)}} s(y-y_0) \right) \right]
    \nonumber \\
    &-\phi_0 \leq\phi\leq \phi_0, \nonumber \\
    &\phi_2(y) = \pi - 2\arctan \left[ \tan\left(\frac{\phi_{0}-\pi}{2}\right)\tanh\left(
    \frac{|\sin \phi_{0}|}{2\sqrt{\cos(\phi_0)}} s(y-y_0) \right) \right] \nonumber \\
    &\phi_0 \leq\phi\leq 2\pi-\phi_0.
\end{eqnarray}
Both solutions evolve between the two equilibrium phases and are shown in Fig.~\ref{mass-fig-T-S}a. Recalling that a change of $2\pi$ in the phase is
associated to a flux quantum $h/(2e)$, we 
see that each of these solitons is associated to a fraction of the flux quantum. 

Interestingly, the spatial dependence of the phase in these two solutions of the double sine-Gordon equation for the junction introduces important
changes in the mass term of Eq.~(\ref{eq-action_0c}). In fact, 
substituting the functions $\phi_j(y)$ in \JS{$m_2(\phi)$}, we find its dependence with
the position along the junction. The results are shown in Fig.~\ref{mass-fig-T-S}b.

\subsection{TRITOPS-TRITOPS junction}
\begin{figure}[t]
\centering
\includegraphics[width=\linewidth]{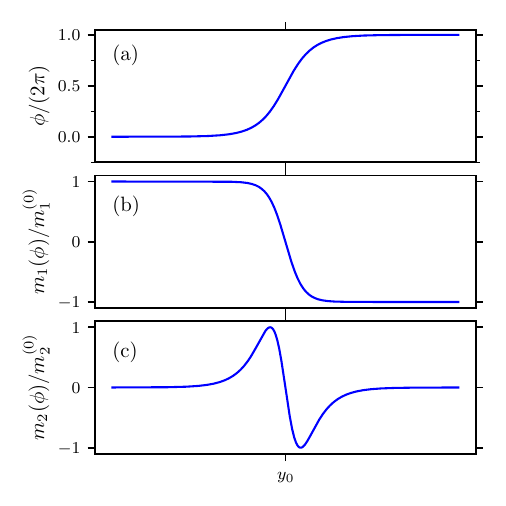}
\caption{Profile for a soliton of the phase (a) and the \JS{ masses $m_{1}(\phi)$  (b) and  $m_{2}(\phi)$ (c), respectively,} for the TRITOPS-TRITOPS junction with $\phi_0=0.12\times 2\pi$.}\label{fig:prof-TT}
\end{figure}

 The equilibrium phase in this case is $\phi_{\rm eq}=0$. As before, we approximate Eq.~(\ref{ephi}) by substituting the log by a constant, which results in 
$E(\phi)\simeq E_{\rm J}-E_0/2-(E_{\rm J}+E_0/2)\cos(\phi)$. 
%{\color{blue}AS: Should we rewrite it as $E(\phi)\simeq(E_{\rm J}-E_0/2)-(E_{\rm J}+E_0/2)\cos(\phi)$? This would explain why we get the perfect sine-Gordon.}{\color{red}GRR: It sounds good to me.}
Therefore $\delta\epsilon(\phi)=\frac{E(\phi)-E(0)}{E_{\rm J}} \simeq \left(1 +\frac{E_0}{2 E_{\rm J}}\right)\left(1-\cos(\phi)\right)$ and
Eq.~(\ref{sol-prof}) reads
\begin{equation}
\int_\pi^{\phi} d \varphi \frac{1}{\sqrt{2 \delta \epsilon(\varphi)}}= y-y_0,
\end{equation}
being $\phi(y_0)=\pi$. The solution for a kink is
\begin{equation}\label{phiT-T}
\phi(y)=4 \arctan\left(e^{\sqrt{\frac{E_0}{2E_{\rm J}}+1}(y-y_0)}\right),
\end{equation}
and it is shown in Fig.~\ref{fig:prof-TT}a. We see that this kink solution is very similar to Eq.~(\ref{kink}) but with a soliton width given by $a^{-1}=\left(\frac{E_0}{2E_{\rm J}}+1\right)^{-1/2}$. 

Substituting in the mass from Eq.~(\ref{intro_TRITOPS-TRITOPS}), we find the following behavior for its variation as a function of the position,
\JS{\begin{eqnarray}\label{mastt}
m_1(y)&=&-m^{(0)}_1 \tanh \left(a\left(y-y_0\right)\right),  \\
m_2(y)&=&-2m^{(0)}_2 \tanh \left(a\left(y-y_0\right)\right)\sqrt{1-\tanh^2\left(a\left(y-y_{0}\right)\right)}. \nonumber
\end{eqnarray}
For an antisoliton, the overall minus sign for  both masses is replaced by a plus sign. The masses $m_{1,2}(y)$ as function of position  are shown in Fig.~\ref{fig:prof-TT}.  Both masses are odd under $y\rightarrow -y$, i.e. $m_{1,2}(-y)=-m_{1,2}(y)$. While $m_1(y\rightarrow \pm \infty)\rightarrow \mp m^{(0)}_1 $, the mass $m_2(y)$ has support only near the soliton, i.e. $m_2(y\rightarrow \pm \infty)\rightarrow 0 $. }

\section{Electron bound states at the fluxon}\label{sec:bound}
We now analyze the coupling between the phase and  the fermions in the edge states when the junction has trapped fluxons, leading to the solitonic profiles discussed in the previous section, \JS{using the Hamiltonian in Eq.~\eqref{eq-action_0c},}
%We start by expressing the Hamiltonian of Eq.~(\ref{heff-class}) in real space as follows
%\begin{equation}\label{heffx}
%H_{\rm eff} = \frac{1}{2}\int dy \; \eta^{\dagger}(y)\left[ m(y) \alpha_0 -i \hbar {\rm v} \alpha_1 \partial_y \right]\eta(y),
%\end{equation}
where we recall that $\eta(y)$ are spinors collecting the members of the Kramers pair of the counter-propagating Majorana modes. 

For the
TRITOPS-S junction there is a single pair of Majorana modes (see Fig.~\ref{fig:sketch}) and we have
%\begin{eqnarray}\label{eq:basis_TRITOPS-S}
%\eta(y)&=&\left(\eta_{\uparrow}(y), \eta_{\downarrow}(y) \right)^T, \\
%& & \alpha_0=\sigma^y,\; \alpha_1=\sigma^z,\;\;\;{\rm TRITOPS-S}.
%\nonumber
%\end{eqnarray}
\JS{\begin{eqnarray}\label{eq:basis_TRITOPS-S}
\eta(y)&=&\left(\eta_{\uparrow}(y), \eta_{\downarrow}(y) \right)^T, \;\;\;{\rm (TRITOPS-S)}.
\end{eqnarray}
Instead, in the TRITOPS-TRITOPS junction this spinor has four components corresponding to the two Kramers pairs associated to the two TRITOPS (labeled by $j=1,2$),
and we have
%\begin{eqnarray}\label{basis0}\label{eq:basis_TRITOPS-TRITOPS}
%\eta(y)&=&\left(\eta_{1,\uparrow}(y), \eta_{1,\downarrow}(y),\eta_{2,\uparrow}(y),\eta_{2,\downarrow}(y)\right)^T, \\
%& & \alpha_0=\tilde{\sigma}^y \; \sigma^z, \;\;\alpha_1=\tilde{\sigma}^z\sigma^z,\;\;\;\;{\rm TRITOPS-TRITOPS},\nonumber
%\end{eqnarray}
\begin{eqnarray}\label{basis0}\label{eq:basis_TRITOPS-TRITOPS}
\eta(y)&=&\left(\eta_{1,\uparrow}(y), \eta_{1,\downarrow}(y),\eta_{2,\uparrow}(y),\eta_{2,\downarrow}(y)\right)^T, \\
& & {\rm (TRITOPS-TRITOPS)}. \nonumber
\end{eqnarray}}
%where $\tilde{\sigma}^j$ and $\sigma^j$ are Pauli matrices acting on 1 or 2 subspace and spin subspace, respectively.
Each of these operators is real, $\eta^{\dagger}(y)=\eta(y)$ as a consequence of the Majorana property $\eta_{k,j,\sigma}^{\dagger}=\eta_{-k,j,\sigma}$ 
being $\eta_{j,\sigma}(y) =\sum_k e^{-iky} \eta_{k,j,\sigma}$ \cite{qi2009time,us}. 
These operators can be expressed in terms of the operators that describe the microscopic properties of the TRITOPS. For  a
two-dimensional system described by a Bogoliubov-de Gennes Hamiltonian in which up and down electrons form $p_x+ip_y$ and $p_x-ip_y$ pairing 
(see $A_{1/2 u}$ and $B_{1/2 u}$ representations of the pairing in Ref.~\onlinecite{us}) they read
\begin{equation}\label{basis-c}
    \eta_{j,\sigma}(y)= \frac{e^{-i s_j s_{\sigma} \pi/4}}{\sqrt{2}} \left( c_{j,\sigma}(y) +i s_j s_{\sigma} c^{\dagger}_{j,\sigma}(y)\right),
\end{equation}
where $s_j =\pm $ for the TRITOPS 1, 2, respectively and $s_{\uparrow}=-s_{\downarrow}=1$. Here, the operators 
$c_{j,\sigma}(y) \;/\; c^{\dagger}_{j,\sigma}(y)$
destroy/create an electron at the coordinate $y$ along the edge in the TRITOPS $j$ with spin $\sigma$. In the TRITOPS-S case we omit the label $j$.

In what follows, we analyze the existence of  bound states by solving the Dirac equation
%\begin{equation}\label{schro}
%\left[ m(y) \alpha_0 -i\hbar {\rm v} \alpha_1\partial_y  \right] \Psi(y)=E \Psi(y),
%\end{equation}
\JS{\begin{equation}\label{schro}
\left[m_{1}\left(y\right)\beta_{1}+m_{2}\left(y\right)\beta_{2}-i\hbar{v}\alpha\partial_{y}\right]\Psi(y)=E\Psi(y),
\end{equation}
with the matrices $\alpha$ and $\beta_{1,2}$ depending on the type of junction and with the mass terms $m_{1,2}(y)$ defined by the soliton profile.}

%We consider configurations of a single soliton as well as soliton-antisoliton. For simplicity, we approximate the dependence of the masses with
%$x$ shown in Figs. \ref{mass-fig-T-S} and \ref{fig:prof-TT} by functions $\mbox{sgn}(x)$. Hence, we consider
%\begin{eqnarray}\label{sol-antisol}
%    m^{\rm s}(x)&=&m_0 \mbox{sgn}(x),\;\;\;\; {\rm soliton} \\
%m^{\rm s-a}(x)&=&m_0-2m_0\left[\theta(x) - \theta(x-L)\right],\;\;\;{\rm soliton-antisoliton}.\nonumber
%\end{eqnarray}
%The problem
%with a soliton-antisoliton profile has solutions describing states localized at the walls situated at $x=0$ and $x=L$.
%For sufficiently close walls the corresponding wave functions overlap and the solutions have 
%$E \neq 0$. These features are generic of the two types of junctions, while the structure of the spinors differ in each case.

\subsection{Single soliton \label{sec_singlesoliton}}
The single soliton case defines the celebrated Jackiw-Rebbi model for a one-dimensional Dirac system with a spatially dependent 
mass term. This problem is known to have a zero mode localized at the point where the mass changes sign. 
For simplicity, we consider $y_0=0$ and we approximate the $y$-dependence of the mass shown in Figs.~\ref{mass-fig-T-S} and \ref{fig:prof-TT} by the function $\mbox{sgn}(y)$. \JS{Given our results of the previous section, we consider $m_{\rm step}=m_2$ in the case of TRITOPS-S and $m_{\rm step}=m_1$ for the TRITOPS-TRITOPS junction, with
\begin{eqnarray}\label{sol-ap}
    m_{\rm step}&=& \pm m_0\,\mbox{sgn}(y), \;\;\;\;\;m_0>0,
\end{eqnarray}
for the antisoliton and soliton, respectively. For simplicity we neglect for now the smaller mass $m_2$ for the TRITOPS-TRITOPS junction.
Below we will include this mass and  avoid the approximation Eq.~\eqref{sol-ap} for  the shape of the mass. } 

\begin{figure}[t]
\centering
\includegraphics[width=\linewidth, trim={0cm 0.5cm 0cm 0.5cm},clip]{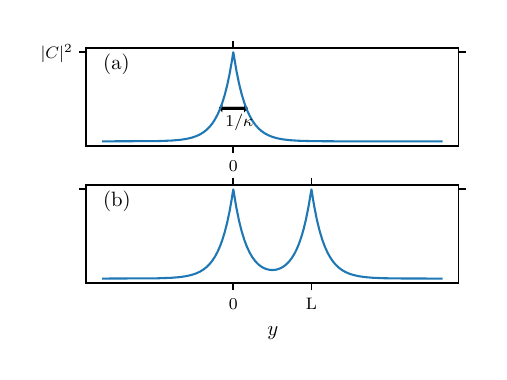}
\caption{(a) Probability density of the bound states of the TRITOPS-S and TRITOPS-TRITOPS junctions described by Eq.~(\ref{eq:localized}). 
(b) Probability density of the bound states of the same junctions for a soliton-antisoliton profile.  }\label{fig:wavef}
\end{figure}

The zero mode is found by solving Eq.~(\ref{schro}) with $E=0$. The result is 
\begin{equation} \label{eq:localized}
    \Psi^{(\pm)}(y) = C \; \Psi_0^{(\pm)} e^{-\kappa|y|}
\end{equation}
with $C$ being a normalization factor, $\kappa=m_0/{\hbar v}$ defines the inverse of the localization length of the zero mode, and $\Psi_0^{(\pm)}$ satisfies $\alpha_0 \alpha_1 \Psi_0^{(\pm)}=\pm i\Psi_0^{(\pm)}$. The spatial dependence of the wave functions is the same for both kinds of junctions and is illustrated in 
Fig.~\ref{fig:wavef} (a).
However, the structure of the spinor is different in the two types of junctions. We now show that this implies a different nature of the bound states.

For the TRITOPS-S junction there is a single solution to Eq.~(\ref{schro}), while the 
TRITOPS-TRITOPS one is doubly degenerate. This reflects the fact that in the latter case time-reversal symmetry is not broken in the equilibrium phase ($\phi_{\rm eq}=0$), hence, there are two localized zero modes that form a Kramers pair. In the TRITOPS-S case, time-reversal symmetry is broken ($\phi_{\rm eq}\neq0$), hence it is possible to have a single solution for a given mass profile. In fact, the solution for a soliton (antisoliton) phase of type $\phi_1$ is equal to the solution for an antisoliton (soliton) in the phase type $\phi_2$.
%\textcolor{purple}{JS: if there is no symmetry breaking at the TRITOPS-S junction, the edge modes are massive. {\color{blue} AS: I guess you mean they are not massive if $\phi=0$. I see now. My mistake. Sorry.} Then one cannot construct normalizable zero modes at the soliton. I did this calculation and checked this.}
The corresponding spinors are
\begin{eqnarray}
{\rm TRITOPS-S}& & \Psi_0^{(\pm)}=(1,\pm 1)^T/\sqrt{2}, \;\;\;\;\;\;  \\
{\rm TRITOPS-TRITOPS}& &
\begin{cases} \label{z-majo-k}
 \Psi_{0,\uparrow}^{(\pm)}=(1,0,\pm 1,0)^T/\sqrt{2}, \\
 \Psi_{0,\downarrow}^{(\pm)}=(0,1,0,\pm 1)^T/\sqrt{2}. \;\;\;\;\;\;
 \end{cases}
\end{eqnarray}
 % Taking into account Eqs. (\ref{eq:basis_TRITOPS-S}), (\ref{eq:basis_TRITOPS-TRITOPS}) and $\gamma_{0\sigma}=\int dy \Psi_{\sigma}^{\dagger}(y) \eta(y)$, these modes can be represented by the following fermionic operators,
 Taking into account Eq.~(\ref{eq:basis_TRITOPS-S}), the resulting Bogoliubov operator in the TRITOPS-S case can be found to be
\begin{align}
{\rm TRITOPS-S} \qquad  \eta^{(\pm)}_0&=\int dy\,\left(\Psi^{(\pm)}(y)\right)^\dagger\eta(y) \notag\\
&=\frac{\tilde{\eta}_{\uparrow}(0)\pm \tilde{\eta}_{\downarrow}(0)}{\sqrt{2}},
%& \textcolor{olive}{= \frac{C}{\sqrt{2}}\int dy\,e^{-\kappa |y|}\,(\eta_{\uparrow}(y)\pm\eta_{\downarrow}(y)).} \label{zero-mode_tritops-s}
\end{align}
where we have defined
\begin{equation}
\tilde{\eta}_\sigma(x)=C\int dy\,e^{-\kappa |y-x|}\eta_{\sigma}(y).
\end{equation}
This zero mode, $\eta_0$, is described by a Majorana operator,
which can be expressed in terms of a spin-1/2 fermionic  operator as follows
\begin{equation}
 \eta_0=
\frac{\left(f_0^{\dagger}+f_0\right)}{\sqrt{2}},    \;\;\;\;\;\;\;\;\;\;\;\;\; f_0=\frac{e^{i\pi/4}}{\sqrt{2}}\left(c_{0,\uparrow}+ic_{0,\downarrow}\right).
\end{equation}

The two degenerate zero modes of the TRITOPS-TRITOPS junction  % of the form $\left(\eta_{1, \sigma} \pm \eta_{2, \sigma}\right)/\sqrt{2},\;\sigma=\uparrow, \downarrow$.
have the same weight at each side of the junction and, using Eq.~(\ref{eq:basis_TRITOPS-TRITOPS}), can be combined to form the two Bogoliubov operators
\begin{align}
&{\rm TRITOPS-TRITOPS}
\begin{cases}\label{zero-modes}
 \gamma_{0,+}=\frac{\tilde{\eta}_{1,\uparrow}(0)\pm\tilde{\eta}_{2,\uparrow}(0) +i\left(\tilde{\eta}_{1,\downarrow}(0)\pm \tilde{\eta}_{2,\downarrow}(0) \right)}{2}, \\
 \gamma_{0,-}=\frac{\tilde{\eta}_{1,\uparrow}(0)\pm \tilde{\eta}_{2,\uparrow}(0)-i\left(\tilde{\eta}_{1,\downarrow}(0)\pm \tilde{\eta}_{2,\downarrow}(0)\right)}{2}.
 \end{cases}
% \\ 
%&\begin{cases}&\textcolor{olive}{\gamma^{(\pm)}_{0,+} = \frac{1}{\sqrt{2}}\int dy\,\left(\Psi^{(\pm)}_{\uparrow} + i\Psi^{(\pm)}_{\downarrow}\right)^\dagger \eta(y)} \\&\qquad \textcolor{olive}{= \frac{C}{2}\int dy\, e^{-\kappa|y|}\left(\eta_{1,\uparrow}(y)\pm\eta_{2,\uparrow}(y) -i\eta_{1,\downarrow}(y)\mp i\eta_{2,\downarrow}(y)  \right)} \\ 
%&\textcolor{olive}{\gamma^{(\pm)}_{0,-} = \frac{1}{\sqrt{2}}\int dy\,\left(\Psi^{(\pm)}_{\uparrow} - i\Psi^{(\pm)}_{\downarrow}\right)^\dagger \eta(y)} \\ &\qquad \textcolor{olive}{= \frac{C}{2}\int dy\,e^{-\kappa|y|}\left(\eta_{1,\uparrow}(y)\pm\eta_{2,\uparrow}(y) +i\eta_{1,\downarrow}(y)\pm i\eta_{2,\downarrow}(y)  \right)}
% \end{cases}
%\LA{ {\rm TRITOPS-TRITOPS}& &
%\begin{cases}
% \gamma_{0,+}=\frac{\left(\gamma_{0,\uparrow}+i \gamma_{0,\downarrow}\right)}{2}=\frac{\left(\gamma_{1,\uparrow}+\gamma_{2,\uparrow}\right)}{2}, \\
% \gamma_{0,-}=\frac{\left(\gamma_{0,\uparrow}-i \gamma_{0,\downarrow}\right)}{2} =-\frac{i\left(\gamma_{1,\downarrow}- \gamma_{2,\downarrow}\right)}{2},
% \end{cases} }
\end{align}
These operators satisfy $\gamma_{0,+}=\gamma_{0,-}^\dagger$, which is typical of zero modes associated to a Kramers pair of Majorana fermions \cite{kwon,camjayi2017fractional,aligia2018entangled}.

Notice that these  zero-energy bound states are very localized. Hence, they are not expected to play a role in the tunneling processes connecting the edge modes 
through the junction, which generates the effective mass $m(\phi)$ \cite{us}. In configurations of many solitons in the junction, they could play some role and $m(\phi)$ should be calculated self-consistently.

\JS{Finally, we discuss the impact of the mass $m_{2}\left(\phi\right)$
on the edge-state spectrum with soliton in  the TRITOPS-TRITOPS junction. As discussed, for a generic
but fixed configuration this mass term breaks time-reversal symmetry.
Hence, one expects the Kramers degeneracy to be lifted. However, as long as both mass terms have a well defined parity under $y\rightarrow -y$ the spectum continues to be doubly degenerate. For the soliton solutions discussed above, both
mass terms are odd under $y\rightarrow-y$. In this case one can identify
two "pseudo-parity" operators 
\begin{eqnarray}
P_{1} & = & \tilde{\sigma}^{0}\sigma^{x}p,\\
P_{2} & = & \tilde{\sigma}^{x}\sigma^{z}p,
\end{eqnarray}
with $pf\left(y\right)=f\left(-y\right)$. 
Both commute with the
 Hamiltonian ${\cal H}$ of
Eq.~\eqref{eq-action_0c}, but do not commute with each other. More generally, $P_1$ and $P_2$ commute with ${\cal H}$ if $\phi\rightarrow 2\pi -\phi$ under $y\rightarrow -y$.
Hence, the spectrum
of ${\cal H}$ with masses that are odd under $y\rightarrow-y$ is doubly degenerate: let $\Psi_1$ be a  simultaneous eigenstate of ${\cal H}$ and $P_1$, then $\Psi_2=P_2 \Psi_1 $ is a distinct eigenstate with same energy; the only alternative would be $\Psi_2=0$, which is not allowed since $P_2\Psi_2=\Psi_1 $ due to $P_2^2=1$. For $m_2=0$ we have  exactly one pair of Kramers degenerate normalizable zero modes. As charge conjugacy requires eigenstates to occur in pairs of opposite energy, the pair must remain at zero energy, i.e. the pair of zero modes cannot be split by the mass term $m_2 \beta_2$. 
Notice $\Psi_2\neq {\cal T} \Psi_1$, i.e. the two zero modes are not Kramers pairs, but  protected by the fact that both masses are odd in $y$. They do  adiabatically connect to  Kramers pairs
at $m_{2}\rightarrow 0$ though. In Appendix \ref{appendix_zero_energy_states} we explicitly demonstrate the existence of doubly-degenerate normalizable zero modes for given antisymmetric mass profiles $m_1(y)$ and $m_2(y)$, with $m_2(y)$ vanishing for $y\rightarrow \pm \infty$.}

%\textcolor{red}{Finally, we analyze the impact of the perturbation described by Eq. (\ref{pert}) in a TRITOPSS-TRITOPSS junction. We first assume that this perturbation does not affect the solution of the sine-Gordon equation, hence, the dependence of the  phase bias as a function of $y$ remains the same as in Eq. (\ref{phiT-T}). Substituting this function into the $m'(\phi)$ entering Eq. (\ref{pert}) we can analyze the effect of this perturbation
%in the solution of the zero modes of Eq. (\ref{schro}),  
%\begin{equation}
%m'(y) \alpha'_0= -2m'_0 \tanh(y-y_0)\sqrt{1-
 %   \tanh^2(y-y_0)}\tilde{\sigma}^z\sigma^y.
%\end{equation}
%Approximating $m'(y)$ by a Heaviside function, as we have done with $m(y)$, 
%We can verify that the solution of the Schr\"odinger equation has two degenerate zero modes of the form
%\begin{eqnarray} \label{eq:localized}
%   \Psi^{(\pm)}_{1}(y) &=& C \; \Psi_{0,1}^{(\pm)} \left( \cosh(y-y_0) \right)^{-\frac{m_0}{\hbar \rm v}} e^{\frac{2 m_0'}{\hbar \rm v}\sech(y-y_0)} \nonumber \\
%    \Psi^{(\pm)}_{2}(y) &=& C \; \Psi_{0,2}^{(\pm)} \left( \cosh(y-y_0) \right)^{-\frac{m_0}{\hbar \rm v}} e^{-\frac{2 m_0'}{\hbar \rm v}\sech(y-y_0)}
%\end{eqnarray}
%with $C$ being a normalization factor and
%\begin{eqnarray}
%{\rm TRITOPSS-TRITOPSS}& &
%\begin{cases} \label{z-majo-ss}
 %\Psi_{0,1}^{(\pm)}=(1, \pm1,\pm 1, 1)^T/2, \\
% \Psi_{0,2}^{(\pm)}=(-1,\pm1,\mp 1, 1)^T/2. \;\;\;\;\;\;
% \end{cases}
%\end{eqnarray}
%These two spinors have the structure of a Kramers partners, but the wave functions differ by a factor depending on $m'_0$ as shown in Eq. (\ref{eq:localized}).}

\subsection{Soliton-antisoliton}
Next, we analyze configurations leading to two sign changes in the mass profile. This corresponds to a soliton-antisoliton (kink-antikink) in the case of TRITOPS-TRITOPS. For the TRITOPS-S case, this can be achieved by means of a fractional soliton-antisoliton  or by two fractional kinks completing the total change from $0$ to $2\pi$ in the phase profile (see Fig.~\ref{mass-fig-T-S}).

We consider here the following function for the mass profile, assuming abrupt changes of the mass sign at $y=0$ and $y=L$
%, which corresponds to approximating the mass profile defined by  $\phi(y)$ by ,
\begin{eqnarray}\label{sol-antisol}
m(y)&=&m_0-2m_0\left[\theta(y) - \theta(y-L)\right].
%\nonumber \\
%m(x)&=&-m_0+2m_0\left[\theta(x) - \theta(x-L)\right].
\end{eqnarray}
%It is important to have in mind that such a solution of the sine-Gordon problem is degenerate with the one defined by the change $m(x) \rightarrow -m(x)$.

The new feature in this case is the possibility of hybridization between the bound states of the single soliton when the distance between the  walls 
is comparable with the localization length, $L < 2 \kappa^{-1}$. In such a scenario, we expect the energy of these bound states to be different from zero \cite{aligia2018entangled,grun2023}. Hence, we search for solutions of Eq.~(\ref{schro}) with $E\neq 0$.

We now briefly discuss the solution of the bound states with finite energy in the case of the TRITOPS-S junction. To this end, we consider Eq.~(\ref{schro}) with $\alpha=\sigma^z$, $\beta_2=\sigma^y$ and
%\begin{equation}\label{eq:antisoliton_soliton_T_T}
%\left[ m(x) \sigma_y -i {\rm v} \sigma_z\partial_x  \right] \Psi(x)=E\Psi(x),
%\end{equation}
 $m_2(y)$ being the mass profile of Eq.~(\ref{sol-antisol}).
 %and $\Psi=(\psi_1, \psi_2)^T$
  It is convenient to perform the rotation
  %a rotation of $-\pi/2$ around $y$, 
$U=e^{i\pi/4 \sigma_y}=\left(\sigma_0+i \sigma^y\right)/\sqrt{2}$
so that 
%the spinor is transformed to $\tilde{\psi}=U\psi$ and 
Eq.~(\ref{schro}) is transformed to
\begin{equation}
\left[i \hbar{v}\partial_y \sigma^x + m(y) \sigma^y\right]\tilde{\Psi} = E \tilde{\Psi},
\end{equation}
where $\tilde{\Psi}=U\Psi=(\tilde{\psi}_1, \tilde{\psi}_2)^T$. After some algebra, we get
\begin{eqnarray}\label{eq:T_S_Schrodinger}
& & \tilde{\psi}_{2} = \frac{i}{E}\left[\hbar{v}\partial_y + m(y)  \right]\tilde{\psi}_{1}, \nonumber \\
& & \left[m^2(y)-\hbar^2{v}^2 \partial_y^2 - {\hbar v} m^{\prime}(y)  \right]\tilde{\psi}_{1}= E^2 \tilde{\psi}_{1}.
\end{eqnarray}
Given the profile of Eq.~(\ref{sol-antisol}), the solutions are
\begin{eqnarray}\label{eq:T_T_solution}
    \tilde{\psi}_{1}(y)&=& \left\{
\begin{array}{ll}
  &-C \chi e^{\kappa(y-L)} , \;\;\;\;\;y\leq 0, \\
   &C \left[-(\chi+1) e^{\kappa(y-L)}+e^{-\kappa(y+L)}\right], \;\;\;\;\;0<y\leq L,  \\
   &C \left[-(\chi+1) e^{\kappa L} + e^{-\kappa L}\right]e^{-\kappa y}, \;\;\;\;y>L,
\end{array}
\right.
\nonumber \\
 \tilde{\psi}_{2,\pm}(y)&=& \pm i \tilde{\psi}_{1}(-y+L),
\end{eqnarray}
with $\chi=\kappa {\hbar v}/m_0$ and
\begin{equation}\label{C-k-ak}
C=\sqrt{\frac{\kappa}{2\left[\chi(\chi+1)-e^{-2\kappa L}(\chi+(\chi+1)\kappa L)\right]}}.
\end{equation}
The energies of the bound states as well as the localization length $\kappa^{-1}$ of these modes are given by
$E_{\pm}=\pm E(L)$ with 
\begin{equation} \label{en-el-s-as}
E(L) = m_0 e^{-\kappa L}, \;\;\;\;\; \hbar^2{v}^2\kappa^2=m_0^2-E^2.
\end{equation}
Notice that $|\chi| \leq 1$ and it can be verified that the solution is always normalizable.
Examples of the spatial dependence of the square of the absolute value of the wave function are shown in Fig.~\ref{fig:wavef} (b).

The spinors in the original basis are
\begin{eqnarray}{\label{eq:Double_soliton}}
\Psi_{\pm} (y) &=& \frac{1}{2}
\left(\begin{matrix}
\tilde{\psi}_{1}(y)\mp i \tilde{\psi}_{1}(-y+L) \\
 \tilde{\psi}_{1}(y)\pm i \tilde{\psi}_{1}(-y+L).
\end{matrix}\right)
\end{eqnarray}
In terms of operators these bound states can be expressed as follows,

\begin{align}
\Gamma=& \frac{1}{\sqrt{2}}\int dy \left[\tilde{\psi}_{1}(y)\eta_{+}(y)+i\tilde{\psi}_{1}(-y+L)\eta_{-}(y)\right]
\end{align}

%\begin{align}
%\Gamma=& \int dy \left(\Psi_{+}(y)\right)^\dagger\eta(y) \\\notag
%=&\frac{1}{2}\int dy \left[\left(\tilde{\psi}_{1}(y)+ i \tilde{\psi}_{1}(-y+L)\right)\eta_{\uparrow}(y)\right.\\\notag
%+&\left.\left(\tilde{\psi}_{1}(y)- i \tilde{\psi}_{1}(-y+L)\right)\eta_{\downarrow}(y)\right]\\\notag
%=&\frac{1}{\sqrt{2}}\int dy \left[\tilde{\psi}_{1}(y)\eta_{+}(y)+i\tilde{\psi}_{1}(-y+L)\eta_{-}(y)\right]
%\end{align}

where $\eta_{\pm}(y)=\frac{1}{\sqrt{2}}\left(\eta_{\uparrow}(y)\pm\eta_{\downarrow}(y)\right)$, while
$\Gamma^{\dagger}, \Gamma$ is a Bogoliubov operator that creates/anhihilates a fermionic excitation with energy $E_+$ given by Eq.~(\ref{en-el-s-as}).

The TRITOPS-TRITOPS case is similar, but with two degenerate states corresponding to combinations of 
the zero-modes localized at $y=0,L$, with the energy given by Eq.~(\ref{en-el-s-as}).
These degenerate bound states for positive energy can be described with the following Bogoliubov operators,
%\begin{eqnarray} \label{bogo}
%\Gamma_{1,+}&=&\frac{1}{2} \left[(\eta_\uparrow(0)+\eta_\downarrow(0)) - i(\eta_\uparrow(L)-\eta_\downarrow(L))  \right],\nonumber \\
%\Gamma_{2,+}&=&\frac{1}{2} \left[(\eta_\uparrow(0)-\eta_\downarrow(0)) + i(\eta_\uparrow(L)+\eta_\downarrow(L))  \right],
%\end{eqnarray}

\begin{eqnarray} \label{bogo}
\Gamma_{\uparrow,+}&=&\frac{1}{\sqrt{2}}\int dy \left[\tilde{\psi}_{1}(y)\eta_{\uparrow,+}(y)+i\tilde{\psi}_{1}(-y+L)\eta_{\uparrow,-}(y)\right],\nonumber \\
\Gamma_{\downarrow,+}&=&\frac{1}{\sqrt{2}}\int dy \left[\tilde{\psi}_{1}(y)\eta_{\downarrow,+}(y)+i\tilde{\psi}_{1}(-y+L)\eta_{\downarrow,-}(y)\right],\\
\end{eqnarray}
where $\eta_{\sigma,\pm}(y)=\frac{1}{\sqrt{2}}\left(\eta_{\sigma,\uparrow}(y)\pm\eta_{\sigma,\downarrow}(y)\right)$,
while those corresponding to the negative-energy states are described with 
\begin{equation}\label{rel-gam}
\Gamma_{\sigma,-}=\Gamma_{\overline{\sigma},+}^\dagger, \;\;\;\;\;\;\sigma=\uparrow, \downarrow,
\end{equation}
with $\overline{\uparrow}=\downarrow$ and  $\overline{\downarrow}=\uparrow$.

%\LA{CREO QUE LO CORRECTO (a menos de fases)}
%\begin{eqnarray} \label{bogo1}
%\Gamma_{1,\sigma,+}&=&\frac{1}{2} \left[\gamma_\sigma(0) + \gamma_\sigma(L)  \right],\nonumber \\
%\Gamma_{2,\sigma,+}&=&\frac{1}{2} \left[\gamma_\sigma(0) - \gamma_\sigma(L)  \right],
%\end{eqnarray}
%SIENDO LOS OPS $\gamma_{0,\sigma},\; \gamma_{L,\sigma}$ los definidos anteriormente.

\subsection{Numerical benchmark}
We now compare the solutions obtained on the basis of the effective Hamiltonian %for the junction defined in Eq.~(\ref{heffx}) 
with numerical results calculated with the Hamiltonians described in Appendix \ref{aplat}, which are defined in finite lattices with $N_x=300$ sites in the $x$-direction  and $N_y=300$ sites along the junction. In the present case we consider open boundary conditions along both directions.

 %Following Ref.~\cite{us} we can define the relation between the parameters of the Hamiltonian of Eq.~(\ref{heff}) and the parameters of the
 %lattice Hamiltonian of  Eq.~(\ref{hfull}) as follows
% \begin{eqnarray}
% {\rm v} & \equiv & \Delta, \\
%     m_0& \equiv &t_{\rm J}, \;\;\;\;\;\;\;\;\;\; \mbox{TRITOPS-TRITOPS}, \nonumber \\
%     m_0& \equiv &|t_{\rm J}|^2/\Delta, \;\;\; \mbox{TRITOPS-S}.\nonumber 
% \end{eqnarray}

 We diagonalize the Hamiltonian in Eq.~(\ref{hfull}) and calculate the
local particle density at the topological side of the TRITOPS-S
junction corresponding to the two states $|\pm\rangle$, which are 
identified as those associated to the  bound states at the kink and antikink, with energy $E_{\pm}$, 
\begin{equation}
\rho^{\pm}_{N_x,\ell}= \sum_{\sigma=\uparrow,\downarrow}\langle \pm|c^{\dagger}_{N_x,\ell,\sigma} c_{N_x,\ell,\sigma}|\pm\rangle = \frac{1}{2}
\langle s|{\bf c}^{\dagger}_{N_x,\ell} \tau^z {\bf c}_{N_x,\ell}|s\rangle,
\end{equation}
being ${\bf c}_{N_x,\ell}=(c_{N_x,\ell\uparrow}, c_{N_x,\ell,\downarrow}, c^{\dagger}_{N_x,\ell,\downarrow}, -c^\dagger_{N_x,\ell,\uparrow})^T$.

\begin{figure}[t]
\centering
\includegraphics[width=\linewidth, trim={0cm 0cm 0cm 0cm},clip]{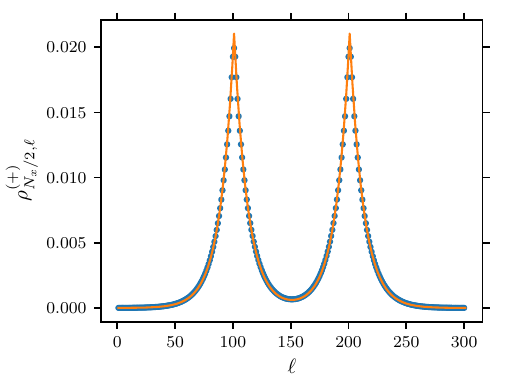}
\caption{Local particle density $\rho^{+}_{N_x,l}$ plotted against the sites of the topological edge of the junction calculated with respect to an $E_{+}$ localized mode % with $m_0/{\rm v}=1/5$.
for a kink-antikink profile $\phi_\ell$ in a ${\rm S}_1-{\rm S}_2 \equiv$ TRITOPS-S junction with $\phi_0=0.12\times\pi$. Parameters are $\mu=-2w$, $t_J=\Delta_p^{(1)}=\Delta_s^{(2)}=4\Delta_s^{(1)}=w/5$, $N_1=100$, $N_2=200$  and $N_x=N_y=300$. The line corresponds to Eq.~(\ref{eq:Double_soliton}) with $m_0=0.0037$ and $v=0.088$.}\label{fig:density}
\end{figure}

Results for the local particle density $\rho^{(+)}_{\ell,N_x}$ illustrating the case of a kink-antikink profile in a TRITOPS-S junction are shown in Fig.~\ref{fig:density}. These  are calculated with 
the following profile for the phase bias:
\begin{equation}\label{phisol-asol}
\phi_\ell=\left\{
\begin{array}{ll}
  &\phi_0
  %+2(i-1)\phi_0
  , \;\;\;\;\;\ell=1, \ldots, N_1-1, N_2+1, \ldots, N_y,  \\
   &\pi
   %(i-1)
   , \;\;\;\;\;\ell=N_1,\; \ell= N_2  \\
   &2\pi-\phi_0
   %+2(i-1)(\pi-\phi_0)
   , \;\;\;\;\ell=N_1,\ldots,N_2,
\end{array}
\right.
\end{equation}
where 
%$i=1,2$ correspond to the two types of solution $\phi_i$ and in addition 
$N_1$ and $N_2$ are the positions of the kink and anti-kink respectively.
The results  of Fig.~\ref{fig:density} correspond to 
%a kink-antikink phase profile $\tilde{\phi}_\ell^{(1)}$ in a TRITOPS-S junction with an equilibrium phase 
$\phi_0=0.12\times\pi$. The energy of this bound state is very small ($\sim 10 ^{-4}$), in consistency with the limit $L=N_2-N_1=100\gg1$ of Eq.~(\ref{en-el-s-as}). In the figure we also show a fit with a function of the form given by Eq.~(\ref{eq:Double_soliton}), indicating the considered values of $m_0$ and
${v}$, while $\kappa$ has been calculated by solving Eq.~(\ref{en-el-s-as}).

\begin{figure}[t]
\centering
\includegraphics[width=\linewidth, trim={0cm 0.5cm 0cm 0.5cm},clip]{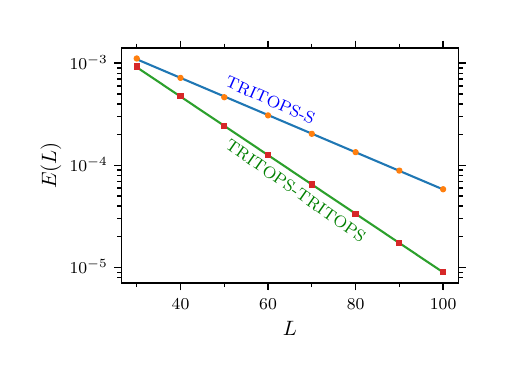}
\caption{TRITOPS-S: Energy due to the hybridization between solitons as function of the distance $L$ between solitons with $\mu=-2w$, $t_J=\Delta_p^{(1)}=\Delta_s^{(2)}=4\Delta_s^{(1))}=t/5$ and $N_x=N_y=300$. TRITOPS-TRITOPS: same with $\mu=-2w$, $5t_J=\Delta_p^{(1)}=t/5$, $N_x=150$, $N_y=300$. Assuming $\kappa\approx m_0/{\hbar v}$, the linear fit corresponds to $\log(E)=-\frac{m_0}{\hbar v} L + \log(m_0)$ obtaining $m_0=0.0037$ and ${\hbar v}=0.088$ for the TRITOPS-S case and $m_0=0.0067$ and ${\hbar v}=0.102$ for TRITOPS-TRITOPS.}
\label{fig:x cubed graph}
\end{figure}

As the distance between solitons gets shorter, the energy increases exponentially following Eq.~(\ref{en-el-s-as}). This behavior
is illustrated 
in Fig.~\ref{fig:x cubed graph} for the two types of junctions and considering the profile of Eq.~(\ref{phisol-asol})
with $\phi_0=0.12\times\pi$ in the case of the TRITOPS-S junction and $\phi_0=0$ for the TRITOPS-TRITOPS one.
Assuming that the separation between kink and antikink is still large, one can approximate $\kappa\approx m_0/v$ in Eq.~(\ref{en-el-s-as}). From there we can get an estimation of the mass $m_0$ and the velocity $v$ as indicated in the caption of Fig.~\ref{fig:x cubed graph}. We recall that the relation between the parameters of the effective continuum model and the lattice Hamiltonian is 
as follows: $m_0 < |t_J|^2/\Delta_{\rm eff}$ for the TRITOPS-S junction and $m_0 < |t_J| $ for the TRITOPS-TRITOPS one, while 
the velocity of the edge modes is $v < \Delta_{\rm eff}$ \cite{us}. Therefore, we find a perfect consistency between the 
description of the bound states in terms of the effective continuum model introduced in Section \ref{sec:bound} and the exact 
numerical results.

%\begin{figure}[H]
%\centering
%\includegraphics[width=\linewidth, trim={0.8cm 10cm 2cm 5cm},clip]{Spectrum_TRITOPS_TRITOPS_with_inset.pdf}
%\caption{TRITOPS-TRITOPS:Energy due to the hybridization between solitons as function of the distance $L$ between solitons with $\mu=-2t$, $5t_J=\Delta_p^{TRITOPS}=t/5$ and $L_x=L_y=150$. Assuming $\kappa\approx m_0^{eff}/v^{eff}$, the linear fit corresponds to $log(E)=-\frac{m_0^{eff}}{v^{eff}} L + log(m_0^{eff})$ obtaining $m_0^{eff}\approx m_0/3$ and $v^{eff}\approx v/2$.}
%\label{fig:x cubed graph}
%\end{figure}

\section{Dynamics of the soliton-antisoliton with the bound states}\label{dyn}
In the previous sections, the focus was on the description of the static problem of the phase bias along the junction defined by the effect of the
CPR, as well as the description of the electron bound states enabled by the phase profile. The aim of this section is to discuss the expected dynamics 
for the coupled system. 

The problem of kink-antikink collisions vs the formation of an oscillating breather bound state in the context of the usual sine-Gordon equation has
a long story \cite{campbell1986kink}. The different scenarios can be analyzed by investigating the time-dependent energy density of the Hamiltonian leading  to Eq.~(\ref{sine-g}),
\begin{equation}\label{junc}
H_{\phi} = \int dy \left[\frac{1}{2} \phi_t^2+\frac{1}{2} \phi_y^2 + \epsilon(\phi)\right],
\end{equation}
with 
\begin{equation}\label{esG}
\epsilon(\phi)\equiv \epsilon_{\rm sG}=1-\cos(\phi).
\end{equation}
This equation is valid for the TRITOPS-TRITOPS junction. For the 
 integrable case of the TRITOPS-S junction, we have Eq.~(\ref{TS}) instead of Eq. (\ref{esG}).
 In what follows we focus on the TRITOPS-TRITOPS case, and we rely on known  analytical results of the dynamics of 
  the usual sine-Gordon model \cite{soliton1}. The conclusions will be similar  
for the other junction.

\begin{figure}[t]
\centering
\includegraphics[width=\linewidth, trim={0cm 0.5cm 0cm 1cm},clip]{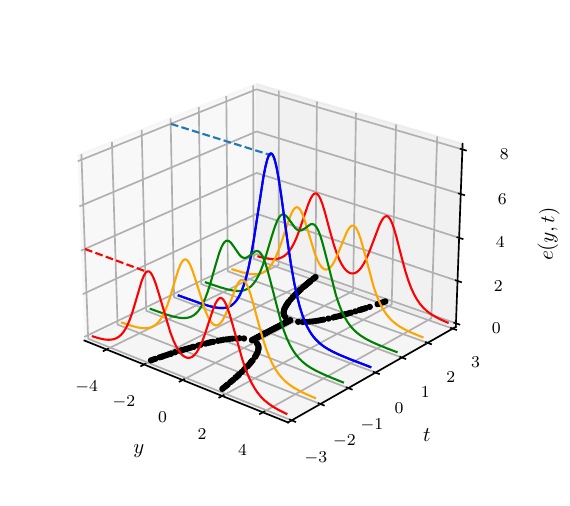}
\caption{Energy density for a collision between a kink and an anti-kink pair with initial velocity $V_k$=0.1 with $\kappa=m_0/{\hbar v}=1$. Black lines correspond to soliton trajectories corresponding to the maximum of the energy density at each given time. Notice that energy is conserved.}
\label{fig:Collision}
\end{figure}

In the framework of the usual sine Gordon Hamiltonian,  kink and antikink solutions travelling in opposite directions with initial velocities $\pm V_k$ are described by the following expression
\begin{equation}\label{sg-k-ak}
    \phi_{\rm sG}(y, t) = 4 \arctan \left[ \frac{V_k \cosh(\delta_k y)}{\sinh(\delta_k V_k t)}\right],
\end{equation}
with $\delta_k=1/\sqrt{1-V_k^2}$. The energy density corresponding to the traveling solitons without taking into account 
the fermionic states is
\begin{equation}\label{en-k-ak}
   e(y,t)= \frac{1}{2} \phi_t^2+\frac{1}{2} \phi_y^2 + \epsilon(\phi).
\end{equation}
The behavior of this function is illustrated in Fig.~\ref{fig:Collision} for a given value of the initial velocity. The total energy is conserved. In particular,
we  see that
adding the maxima of $e(y,t)$ for the two separate solitons we get the maximum at the collision, reflecting the fact that the collision is perfectly elastic. The trajectory 
defined by these maxima as  functions of time is plotted in the horizontal plane.  The kink and anti-kink preserve their shape thereafter and as they approach, they  accelerate. The result is a shift in the trajectories of each soliton with respect to the trajectory without collision \cite{campbell1986kink}.

The aim of the present section is to discuss the behavior of the joint time-dependent energy density corresponding to the Hamiltonian (\ref{junc}) for the phase as well the time-dependent energy of the bound states. We consider the case of a slow motion of the phase profile, so that we can describe the dynamics of the
hybridized 
fermionic modes as a sequence of snapshots where the states have energy $E\left(L(t)\right)$ given by Eq.~(\ref{en-el-s-as}). 

\begin{figure}[t]
\centering
\includegraphics[width=\linewidth, trim={0cm 0.4cm 0cm 0.5cm},clip]{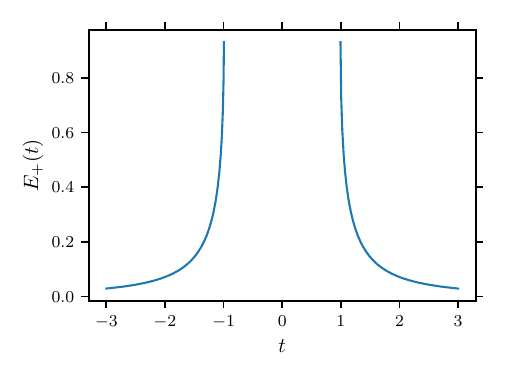}
\caption{Energy as a function of the distance of the fermionic bound state repulsion, being $\kappa=m_0/\hbar v=1$.}
\label{fig:Repulsion}
\end{figure}

When the soliton and antisoliton are far apart, the electron states have zero energy, hence, we expect that their propagation takes place following the description of Eq.~(\ref{en-k-ak}).
However, as they approach, the bound states hybridize.  
%Hence, the total energy 
%density as a function of time reads 
%\begin{equation}
%   e(x,t)= \frac{1}{2} \phi_\tau^2+\frac{1}{2} \phi_z^2 + \epsilon(\phi) + E(L),
%\end{equation}
 We
calculate $L(t)$ from the distance between the maxima of the energy density as a function of time predicted by
Eq.~(\ref{en-k-ak}).  This approach is valid as long the soliton shape is well defined (before and after the collision).
For sufficiently large $L$, we can approximate Eq.~(\ref{en-el-s-as}) by $E_{\pm}(L)\approx \pm m_0 e^{-\frac{m_0}{\hbar v}L}$.
In this regime, we express the Hamiltonian for the subgap fermionic modes  as follows
\begin{equation}
H_f(L) = \frac{1}{2} \sum_{s = \pm } \sum_{\sigma=\uparrow,\downarrow} E_{s}(L) \Gamma^{\dagger}_{\sigma,s} \Gamma_{\sigma,s},
\end{equation}
with $\Gamma_{\sigma,s}, \; \sigma=\uparrow,\downarrow,\;s=\pm $ given by Eq.~(\ref{bogo}). Using the relations obeyed by the operators describing the positive and negative-energy states, Eq.~(\ref{rel-gam}), we can also express this Hamiltonian  as follows
\begin{equation}
H_f(L) =  E_{+}(L) \sum_{\sigma}  \Gamma^{\dagger}_{\sigma,+} \Gamma_{\sigma,+}-E_{+}(L).
\end{equation}
The extra contribution to the total energy originated in the subgap fermionic states depends on the nature and the evolution 
of the many-body state. If the zero modes of the large-$L$ configuration are occupied by two fermions, the traveling
kink and anti-kink experience 
an effective attraction of amplitude $- E_{+}(L)$ as $L$ overcomes the critical length $\xi_c \simeq \hbar {v}/m_0$. Instead, if the
zero modes are occupied by four fermions in the separated solitons, they experience a repulsion of amplitude
$E_{+}(L)$ as they approach. Furthermore, it is possible to have
a change in the parity of the ground state because excitations of single quasiparticles near defects. In such a case, it is also possible to have a single occupied
state with positive energy, and the fermionic states are neutral regarding the soliton dynamics. 
The behavior of   $E_{+}(L(t))$  as a function of time is shown in Fig.~\ref{fig:Repulsion}. Interestingly, a very similar discussion was 
earlier presented in the context of fermionic energies in a multi-soliton background \cite{Altschul}. 

In the case of the TRITOPS-S junction, we can follow a similar reasoning, by assuming that the soliton-antisoliton dynamics is also
described by Eq.~(\ref{sg-k-ak}) and by the following Hamiltonian
\begin{equation}
H_f(L) =  E_{+}(L) \Gamma^{\dagger}_{+} \Gamma_{+}-E_{+}(L)/2.
\end{equation} In this junction,
 the two Majorana fermions localized at the kink and at the antikink combine in a fermionic mode as they approach one another. This results in an attraction 
equal to $-E_{+}(L)/2$ 
%{\color{blue}AS: I believe it is $-E_{+}(L)/2$. Should we give the formula like $H_f(L) =  E_{+}(L) \Gamma^{\dagger}_{+} \Gamma_{+}-E_{+}(L)/2$?} 
In the presence of  poisoning corresponding to  a change in the parity of the many-body ground state, the positive-energy state is occupied, which generates a repulsion 
of amplitude $E_{+}(L)/2$.
%{\color{blue}AS: again $1/2$}.

\section{Summary and Conclusions}
We have analyzed the joint dynamics of the phase bias $\phi$ and the Majorana edge modes at the different types of Josephson junctions formed with two dimensional (2D) time-reversal invariant topological superconductors (TRITOPS). In particular, we have considered TRITOPS-TRITOPS and
 junctions between a TRITOPS and a non-topological  superconductor (TRITOPS-S).  The analysis is mainly based on effective models for the
 junctions, which consist of Dirac Hamiltonians with mass depending on the phase bias and the results are contrasted against the numerical solutions of lattice models.
We have extended the study of Ref. \onlinecite{gros-stern-pnas}, which focused on junctions of 2D topological superconductors with broken time-reversal symmetry (class D) in two main directions. In the case of the TRITOPS-TRITOPS junction we have considered both, the direct hybridization between the topological edge modes and the hybridization between these modes, and the supra-gap states of the superconductor at the other side of the junction. The latter is a second-order process that generates an extra effective mass, which
 couples the two partners of the edge Kramers' pair of each TRITOPS. In the case of the TRITOPS-S junctions this is the only massive term playing a role.

 We have first analyzed the effect of the phase fluctuations. These may play a critical role in the case of 
TRITOPS-S junctions and may destroy the stability of a state with broken time-reversal symmetry. We have shown that the phase stiffness must 
overcome a critical value for the mean-field description to be stable. 

As a second step, assuming the stability of the mean field description for the phase, we have
analyzed the effect of fluxons in the junction. This defines a sine-Gordon model (for the TRITOPS-TRITOPS) and a double sine-Gordon model
(for the TRITOPS-S case). We have calculated the solitonic profiles of the kink/antikink solutions of these models. These generate changes in the sign of the mass term
of the Dirac Hamiltonians describing the coupling of the Majorana edge states.

Therefore, the third step was to study the impact of the solitonic profiles  on the fermionic states. 
We have shown the existence of zero modes localized at the fluxons. These are Majorana zero modes in the case of a TRITOPS-S junction and 
a Kramers pair of Majorana modes in the case of the TRITOPS-TRITOPS junction. In the latter case, we have shown
that these zero modes remain robust against the effect of the second-order mass term, in spite that the fact that this term breaks time-reversal symmetry. This property is due to extra symmetries of the junction, assuming spin-preserving tunneling processes.  

The existence of such localized zero modes is very interesting in view of potential applications,
since they can be manipulated and moved in these systems by the motions of the fluxons at the junction. 

Finally, we have analyzed the effect of the localized fermionic modes in the collision dynamics of soliton-antisolitons. While such a collision is
perfectly elastic in non-topological junctions, the hybridized zero modes of the junctions studied in this paper originate an effective
atractive or repulsive interaction, depending on the occupation of these modes.
These properties should be very useful for the experimental detection of these states. 

\section{Acknowledgements}
We are grateful to A. Ustinov for many interesting discussions. AS is also grateful to Jay D. Sau and Victor Yakovenko for illuminating discussions. 
LA and GRR thank 
support from CONICET as well as FonCyT, Argentina, through grants  PICT-2018-04536 and PICT 2020-A-03661. We also thank support from
Alexander von Humboldt foundation (LA) and the SPUK collaboration (JS and LA). AR and AS were supported by the DFG under the grant No. SH81/7-1. JS was supported by DFG through Project ER 463/14-1 and acknowledges a Weston Visiting Professorship
at the Weizmann Institute of Science where part of this work was performed. LA would also like to thank the Institut Henri Poincar\'e (UAR 839 CNRS-Sorbonne Universit\'e) and the LabEx CARMIN (ANR-10-LABX-59-01) for their support.

\appendix

\section{Lattice model for the junction}\label{aplat}
We consider the following Hamiltonian
\begin{equation}\label{hfull}
H=\sum_{\ell=1}^{N_y} H_{\ell}, \;\;\;\; \;\;\;\;\; H_{\ell}=\sum_{\alpha={\rm S_1,S_2}} H_{\alpha,\ell} + H_{{\rm J},\ell}.
\end{equation}
The Hamiltonian $H_{\alpha,\ell}$ corresponds to the TRITOPS Hamiltonian expressed in a ribbon with $N_y$ sites along the junction and $N_x$ sites in the other direction. 
The Hamiltonian for the tunneling junction is $H_{\rm J} =\sum_\ell H_{{\rm J},\ell}$, with
\begin{equation}
H_{\mathrm{J},\ell}=t_{\mathrm{J}}\sum_{\sigma}\left( e^{i\phi_\ell /2} c_{\ell,1\sigma}^{\dagger } c_{\ell,2,\sigma}+\text{H.c.}\right) ,  \label{junction}
\end{equation}
where $c_{\ell,1,\sigma}^{\dagger }$ ($c_{\ell,2, \sigma}^{\dagger }$) creates an electron with spin $\sigma$ in the
superconductor $\mathrm{S1}$  ($\mathrm{S2}$) at the site located at the boundary contacting the junction. The phase bias at the junction contains the information of the
soliton profile. 

\begin{figure}[t]
\centering
\includegraphics[width=\linewidth, trim={0cm 0.3cm 0cm 0.3cm},clip]{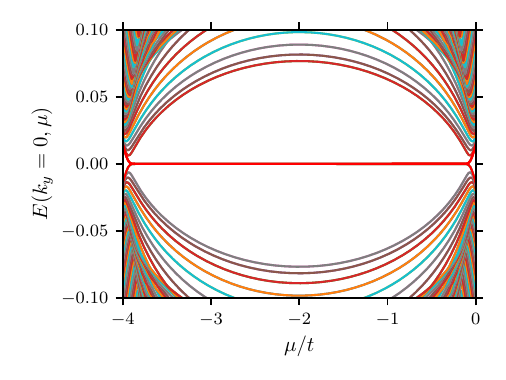}
\caption{Energy for the states with $k_y=0$ vs.\ $\mu$ for a TRITOPS superconductor periodic in the $y$-direction with $\Delta_p=4\Delta_s=w/5$ and $w=1$. The topological phase corresponds to the values of $\mu$ with double-degenerate zero-energy states (depicted in red).}
\label{fig:Topo_phase}
\end{figure}

\begin{figure}[t]
\centering
\includegraphics[width=\linewidth, trim={0cm 0.3cm 0cm 0.3cm},clip]{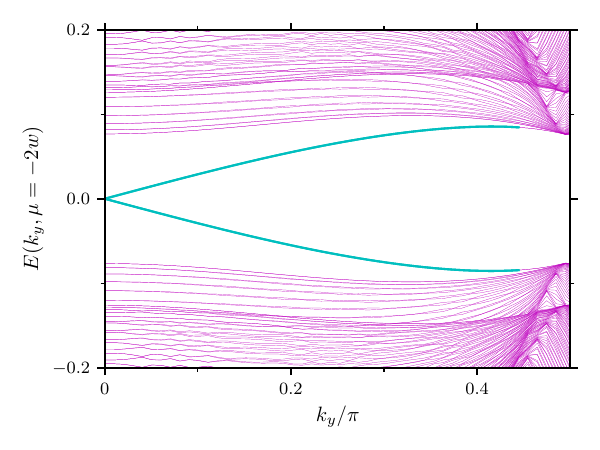}
\caption{Energy for the states with $\mu=-2w$ vs.\ $k_y$ for a TRITOPS superconductor periodic in the $y$-direction with $\Delta_p=4\Delta_s=w/5$ and $w=1$. The topological edge states correspond to the double-degenerate states inside the gap (depicted in light blue).}
\label{fig:Topo_phase_k}
\end{figure}

The Hamiltonian for the TRITOPS formulated in the reciprocal space reads
$H = \frac{1}{2}\sum_{\bf k} {\bf c}^{\dagger}_{\bf k} H_{\bf k}{\bf c}_{\bf k}$, with
${\bf c}_{\bf k}= \left(c_{\bf k, \uparrow},c_{\bf k, \downarrow}, c^{\dagger}_{\bf -k, \downarrow}, - c^{\dagger}_{\bf -k, \uparrow}\right)^T$ and ${\bf k}=\left( k_x,k_y \right)$, while the Bogoliubov de Gennes Hamiltonian matrix reads
\begin{equation}\label{h-p-pm}
    H_{\bf k}  =  \xi_{\bf k} \tau^z S^0 +  \tau^x \boldsymbol{S} \cdot \boldsymbol{\Delta}^{x, y}_{\boldsymbol{k}} +\tau^x S^0 \Delta_s.
\end{equation}
The Pauli matrices $\tau^{x,y,z}$ and $\boldsymbol{S}=(S^{x},S^y,S^z)$ act, respectively,  on the particle-hole and spin degrees of freedom,
while $\tau^0, S^0$ are $2\times 2$ identity matrices. The dispersion relation is defined in terms of a hopping element $w$ as  $\varepsilon_{\bf k}=- 2 w \left(\cos k_x + \cos k_y \right)$, 
hence $\xi_{\bf k}=\varepsilon_{\bf k} -\mu $, being $\mu$ the chemical potential. For simplicity, we consider only nearest neighbor hopping in $\varepsilon_{\bf k}$. 
For the TRITOPS we consider a $p$-wave pairing term defined by
\begin{equation}
\boldsymbol{\Delta}^{x, y}_{\boldsymbol{k}} =  \Delta_p \left(\sin k_x \;{\bf n}^{x}+ \sin k_y \;{\bf n}^{y} \right),
\end{equation}
with 
 ${\bf n}^{x,y}$ being unit vectors along the $x,y$-directions along with a local s-wave component modulated by $\Delta_s$. We choose the values of the parameters corresponding to the topological phase. To verify this property we confirm the existence of Kramers pairs of topological edge states as shown in Fig.~\ref{fig:Topo_phase} and Fig.~\ref{fig:Topo_phase_k}. 
 The non-topological superconductor (S) is modeled with $\Delta_p=0,\; \Delta_s\neq 0$.
 \\
 
\section{Symmetries of the TRITOPS-TRITOPS junction \label{appendix_symmetries}}
Consider the continuum limit of the Hamiltonian in (\ref{h-p-pm}), modelling a TRITOPS with p-wave pairing, which up to leading order in the momenta and with $\Delta_s=0$ reads
\begin{align}
    \mathcal{H}^{(\pm)} = -\mu^{(\pm)}(x)\tau^{z}S^{0}+v\tau^{x}(p_{x} S^{x}+p_{y} S^{y}). \label{eq:continuumBdGHam}
\end{align}
$\mu(x)$ is assumed to change sign at the edge of a given sample and the topological phase corresponds to $\mu/v>0$. Thus, we distinguish $\mu^{(+)}(x\gtrless0)\lessgtr 0$
and $\mu^{(-)}(x\gtrless 0)\gtrless 0$ describing right and left edges, respectively. For $p_y=0$, we find in each case two normalizable zero modes bound to the respective edge satisfying $\mathcal{H}\Phi_0(x) = 0$ given by
\begin{align}
\begin{split}
    \Phi^{(\pm)}_{0,\uparrow}(x) &= \frac{N}{2} e^{\pm \int^x_0 dx^\prime \mu(x^\prime)/v}(1\pm i,0,0,-1\pm i)^{T}, \\
    \Phi^{(\pm)}_{0,\downarrow}(x) &= \frac{N}{2} e^{\pm \int^x_0 dx^\prime \mu(x^\prime)/v}(0,1\mp i,1\pm i,0)^{T}.
    \label{eq:leftrightedgezeromodes}
\end{split}
\end{align}
Note that the solutions are chosen such that they are orthonormal
and are invariant under charge conjugation $\mathcal{C}\Phi_{0,\sigma}^{(\pm)}(0)=\Phi_{0,\sigma}^{(\pm)}(0)$,
where the charge conjugation operator is given by $\mathcal{C}=\tau^{y} S^{y}\mathcal{K}\equiv\mathcal{U}_{C}\mathcal{K}$. Additionally, it is worth noting that the two respective solutions
are Kramer's pairs $\mathcal{T}\Phi_{0,\downarrow}^{(\pm)}=\Phi_{0,\uparrow}^{(\pm)}$,
with the time-reversal operator $\mathcal{T}=iS^{y}\mathcal{K}$. Including finite momenta $p_y$ entails projecting the corresponding term in the Hamiltonian onto this basis, which yields
\begin{equation}
\left(\Phi_{0,\sigma}^{(\tau)}(0)\right)^{\dagger}\tau^{x}S^{y}\Phi_{0,\sigma^{\prime}}^{(\tau^{\prime})}(0)=(\tilde{\sigma}^{z})^{\tau,\tau^{\prime}}(\sigma^{z})^{\sigma,\sigma^{\prime}},
\end{equation}
as given in Eq.~(\ref{intro_TRITOPS-TRITOPS}). 
%{\color{orange} \textbf{AR: Comment on notation:} Here and in Appendix A, the matrices $\sigma$ are not used consistently with the main text.}\\
Transforming the anti-unitary operators $\mathcal{C}$ and $\mathcal{T}$ to this basis, one finds $\mathcal{C}=\mathcal{K}$ and $\mathcal{T}=i\sigma^y\mathcal{K}$.\\

Additionally, we can identify the symmetry corresponding to exchanging the right and the left edge with each other. On the level of the Hamiltonian in
(\ref{eq:continuumBdGHam}), this is achieved by simultaneous
inversion along the $x$-axis $P_{x}$, i.e. $P_{x}f(x)=f(-x)$,
and rotating with $U_{X}=\tau^{z}S^{x}$, 
\begin{equation}
(U_{X}P_{x})\,\mathcal{H}^{(\pm)}\,(P_{x}U_{X})=-\mu^{(\pm)}(-x)\tau^{z}+v\tau^{x}(p_{x}S^{x}+p_{y}S^{y}).
\end{equation}
This is the same Hamiltonian as (\ref{eq:continuumBdGHam}) only
with the direction of the sign change of $\mu$, which determined
whether we called it a right or left edge, reversed. $U_{X}$
connects the right and left edge modes in (\ref{eq:leftrightedgezeromodes})
with each other
\begin{equation}
U_{X}\Phi_{0,\uparrow}^{(+)}=\Phi_{0,\downarrow}^{(-)},\quad U_{X}\Phi_{0,\uparrow}^{(-)}=\Phi_{0,\downarrow}^{(+)}.
\end{equation}
Expressed in the new basis, it therefore holds
\begin{equation}
U_{X}=\tilde{\sigma}^{x}\sigma^{x}.
\end{equation}

\section{Derivation of the fluctuation-corrected equation of state \label{appendix_gapeq}}
In order to derive the fluctuation-corrected 
equation of state in (\ref{gap-eq_fluct}), following Ref.~\onlinecite{reich2023magnetization}, we expand the action in Eq.~(\ref{eq-GNY-action}) up to second order in the fluctuations $\delta\phi$ around some assumed minimum $\braket{\phi}=\tilde{\phi}$ of the ground state energy, $\phi(x,t)=\tilde{\phi}+\delta\phi(x,t)$. After a Wick rotation to Euclidean time and integrating out the fermions, we find the effective action at some temperature $T$ and length of the system $L$ to read
\begin{align}
    S_{\rm eff} = \frac{1}{2}\frac{T}{L}\sum_{q,\omega_m}&\mathcal{D}^{-1}(q,\omega_m)|\delta\phi(q,\omega_m)|^2 \notag\\
    &+ \frac{1}{2}\frac{L}{T}E_{\rm J}\tilde{\phi}^2 - \frac{1}{2}{\rm tr}\log\mathcal{G}_0^{-1}
\end{align}
where
\begin{align}
    \mathcal{D}^{-1}(q,\omega_m) = \hbar K\left(\frac{\omega_m^2}{c_{\rm J}}+c_{\rm J}q^2\right)+E_{\rm J}+\Pi(q,\omega_m),
\end{align}
with 
\begin{align}
    \Pi(q,\omega_m) &= \frac{m_0^2}{4}\frac{T}{L}\sum_{k,\varepsilon_n}{\rm tr}_{2\times 2}\left[\mathcal{G}_0(k,\varepsilon_n)\,\alpha_0\right.\\&\hspace{2cm}\left.\mathcal{G}_0(k+q,\varepsilon_n+\omega_m)\,\alpha_0\right]. \notag
\end{align}
and
\begin{align}
    \mathcal{G}_0^{-1} = \begin{pmatrix} \hbar(\partial_\tau-i{v}\partial_y) & -im_0\tilde\phi \\ im_0\tilde\phi & \hbar(\partial_\tau+i{v}\partial_y) \end{pmatrix}.
\end{align}
$\omega_m$ and $\epsilon_n$ are the bosonic and fermionic Matsubara frequencies respectively. The ground state energy density in terms of $\tilde\phi$ can now be determined from $E(\tilde\phi)=\left(-\frac{T}{L}\log\int\mathcal{D}(\delta\phi)e^{-S_{\rm eff}}\right)_{T\rightarrow 0}$. Differentiating this expression with respect to $\tilde{\phi}^2$ finally results in Eq.~(\ref{gap-eq_fluct}) of the main text, where we used \cite{vaks_collective_1962} that for $T\rightarrow 0$ 
\begin{align}
    \Pi(q,\omega) = \frac{m_0^2}{4\pi\hbar{v}}\left[\frac{1}{2}\log\left(\frac{m_0^2\tilde{\phi}^2}{4\Lambda^2}\right)+\frac{\sqrt{1+r^2}}{r}\rm{Arsinh}(r)\right]
\end{align}
with $r=\hbar\sqrt{\omega^2+v^2q^2}/(2m_0\tilde\phi)$.

\section{Zero-energy solutions to double mass Dirac equation\label{appendix_zero_energy_states}}

The Dirac equation for zero modes in the TRITOPS-TRITOPS junction in the presence of a phase slip soliton reads
\begin{gather}
    \left(-i \hbar v \partial_y\tilde\sigma^z\sigma^z + m_1(y)\tilde\sigma^y\sigma^z + m_2(y)\tilde\sigma^0\sigma^y\right)\psi(y) = 0 \\
    \Leftrightarrow\quad \left(\hbar v \partial_y + m_1(y)\tilde\sigma^x\sigma^0 + m_2(y)\tilde\sigma^z\sigma^x\right)\psi(y) = 0 \label{eq:zeromodeseqinapp}
\end{gather}
where asymptotically $m_1(y\rightarrow\pm\infty)=\pm m_1^{(0)}$ and $m_2(y\rightarrow\pm\infty)=0$ with both mass terms antisymmetric $m_{1,2}(y)=-m_{1,2}(-y)$. Note that $m_2\neq 0$ breaks time-reversal symmetry and one would thus expect the Kramers degeneracy to be lifted in its presence. However, as argued in the main text, this turns out not to be the case if both masses have well-defined parity, due to the existence of the two non-commuting "pseudo-parities" $P_1=\tilde\sigma^0\sigma^x p$ and $P_2=\tilde\sigma^x\sigma^z p$ (see the discussion in Section \ref{sec_singlesoliton}). As a consequence, the two-fold degeneracy of the zero modes bound to the soliton also persists. Here, in order to demonstrate this, we will explicitly derive the two zero-mode solutions to Eq.~(\ref{eq:zeromodeseqinapp}) in the simple case of the soliton profile being given by 
\begin{gather}
    m_1(y) = m_1^{(0)}\text{sgn}(y), \label{eq:mass1profileapp}\\ 
    m_2(y) = m_2^{(0)}\text{sgn}(y)\,\Theta(W/2-|y|).\label{eq:mass2profileapp}
\end{gather}
We start by looking for a solution $\ket{\psi_1}$ to \eqref{eq:zeromodeseqinapp} which is simultaneously an eigenvector of $P_1$ and make the ansatz
\begin{align}
\ket{\psi_1}=\ket{\xi_1}\otimes\ket{+},
\end{align}
where $\sigma_x\ket{\pm}=\pm\ket{\pm}$. Eq.~(\ref{eq:zeromodeseqinapp}) then reduces to
\begin{align}
    \left(i\hbar v \partial_{y}+m_{1}(y)\tilde\sigma_{x}+m_{2}(y)\tilde\sigma_{z}\right)\xi_{1}=0,
\end{align}
to which the normalizable solution can be found to read
\begin{widetext}
\begin{align}
\xi_{1}(y)&\propto\begin{cases}
\begin{pmatrix}1\\
1
\end{pmatrix}e^{m_1^{(0)}(y+W/2)}, & y<-W/2,\\
\begin{pmatrix}1\\
1
\end{pmatrix}\cosh\left(\frac{y+W/2}{\lambda}\right)+\lambda\begin{pmatrix}m_1^{(0)}+m_2^{(0)}\\
m_1^{(0)}-m_2^{(0)}
\end{pmatrix}\sinh\left(\frac{y+W/2}{\lambda}\right), & -W/2<y<0,\\
\begin{pmatrix}1\\
1
\end{pmatrix}\cosh\left(\frac{y-W/2}{\lambda}\right)-\lambda\begin{pmatrix}m_1^{(0)}+m_2^{(0)}\\
m_1^{(0)}-m_2^{(0)}
\end{pmatrix}\sinh\left(\frac{y-W/2}{\lambda}\right), & 0<y<W/2,\\
\begin{pmatrix}1\\
1
\end{pmatrix}e^{-m_1^{(0)}(y-W/2)}, & y>W/2,
\end{cases}
\end{align}
\end{widetext}
where we introduced $\left(m_1^{(0)}\right)^2+\left(m_2^{(0)}\right)^2\equiv\lambda^{-2}$. The second, degenerate zero mode $\ket{\psi_2}$ is then found by application of $P_2$ to the first solution 
\begin{align}
\ket{\psi_2}=P_2\ket{\psi_1}=\tilde\sigma^x\ket{\xi_1}\otimes\ket{-}.
\end{align}
Note that for $m_2^{(0)}\rightarrow 0$, $\ket{\psi_1}$ and $\ket{\psi_2}$ are indeed transformed into one another by the time-reversal operator $\mathcal{T}$. 

It is easy to check that for mass profiles of the form as given in \eqref{eq:mass1profileapp} and \eqref{eq:mass2profileapp}, but with an added asymmetry of some sort, in general no non-trivial solutions to Eq.~\eqref{eq:zeromodeseqinapp} and thus no zero modes exist. Their degeneracy and thus existence is dependent on the parity and with that also sensitive to any impurities or inhomogeneities. 
 \\

\end{document}